\documentclass[11pt, a4paper]{article}
\usepackage{jheppub}
\usepackage{amsmath}

\usepackage{graphicx}
\usepackage{amsfonts}
\usepackage{slashed}
\usepackage{braket}
\usepackage{multirow}
\usepackage{comment}
\usepackage{cases}
\usepackage{bookmark}
\usepackage{subfigure}

\allowdisplaybreaks

\newcommand{\numberthis}{\addtocounter{equation}{1}\tag{\theequation}}

\newcommand{\cB}{{\cal B}}

\newcommand{\cD}{{\cal D}}

\newcommand{\cF}{{\cal F}}
\newcommand{\cG}{{\cal G}}

\newcommand{\cJ}{{\cal J}}

\newcommand{\cL}{{\cal L}}
\newcommand{\cM}{{\cal M}}

\newcommand{\cO}{{\cal O}}
\newcommand{\cP}{{\cal P}}

\newcommand{\cS}{{\cal S}}
\newcommand{\cT}{{\cal T}}

\newcommand{\cZ}{{\cal Z}}

\newcommand{\GT}{{\cG_\cT}}
\newcommand{\FT}{{\cF_\cT}}
\newcommand{\GS}{{\cG_\cS}}
\newcommand{\FS}{{\cF_\cS}}
\newcommand{\dtpsi}{{\dot{\tilde\psi}}}

\title{Non-Gaussianity and Strong-Coupling Problem in a Two-Field DHOST Bouncing Model}
\author{Ok Song An,\,\,}
\emailAdd{os.an@ryongnamsan.edu.kp}
\author{Kon Hong,\,\,}
\author{Jin U Kang,\,\,}
\emailAdd{ju.kang0718@ryongnamsan.edu.kp}
\author{and\, Ui Ri Mun}
\emailAdd{ur.mun0826@ryongnamsan.edu.kp}

\affiliation{Department of Physics, \textbf{Kim Il Sung} University, Ryongnam Dong, Taesong District, Pyongyang, Democratic People's Republic of Korea}

\abstract{
We recently constructed a two-field Degenerate Higher-Order Scalar-Tensor (DHOST) bouncing model which is fully viable at the \textit{linear} level \cite{An:2025xxx}. This model is completely free of Belinski-Khalatnikov-Lifshitz (BKL) instability, ghost instability, gradient instability and superluminality. It also predicts the scalar spectral index and tensor-to-scalar ratio consistent with observations. The aim of this paper is to extend the viability of the model to the \textit{non-linear} level.
	To this end, we first refine the original model such that its prediction on the (local) non-Gaussianity parameter $f_{NL}$ agrees with observations, leaving the viability of the model at the linear level intact.  We furthermore demonstrate that the strong-coupling scale is well above the characteristic background energy scale all the time. Our model indeed exemplifies the fully viable two-field DHOST bouncing model, in the sense that it is weakly-coupled, stable and non-superluminal as well as consistent with observations.}
\preprint{}

\begin{document}
\maketitle
	
\section{Introduction}

Bouncing cosmologies (see \cite{Novello:2008ra,Battefeld:2014uga,Brandenberger:2016vhg} for a review) are regarded as a compelling alternative to the inflationary paradigm \cite{Guth:1980zm,Starobinsky:1980te,Linde:1981mu}, since they provide a natural solution to the initial singularity problem \cite{Penrose:1964wq,Hawking:1970zqf} that the standard Big Bang cosmology and even the inflationary scenarios \cite{Borde:1996pt,Borde:2001nh} suffer from (see however \cite{Lesnefsky:2022fen}). But within the framework of General Relativity cosmic bounce requires violation of the Null Energy Condition (NEC), which generally leads to various pathologies such as ghost instability, gradient instability and superluminality \cite{Rubakov:2014jja,Mironov:2024xxx}. 

Modified gravity theories have been widely employed to obtain  a non-pathological bouncing scenario. Among these, an exceptional subclass \cite{Mironov:2020pqh,Mironov:2024xxx} of Degenerate Higher-Order Scalar-Tensor (DHOST) Ia theories \cite{Langlois:2015cwa,BenAchour:2016cay,Langlois:2017mxy,Langlois:2018dxi,Kobayashi:2019hrl}  deserves a special attention, since any DHOST theory which can realise an observationally compatible bouncing scenario without the aforementioned pathologies  falls into this subclass. However, it was shown in  \cite{An:2025xxx} that the exceptional subclass of DHOST Ia theories serves only as a necessary condition, not a sufficient condition for the non-pathological bounce. Upon deriving the sufficient condition, the authors of \cite{An:2025xxx} constructed an explicit bouncing model for the first time  which admits a cosmological bouncing solution theoretically healthy and phenomenologically viable \emph{in the linear regime}, namely
\begin{itemize}
	\item the model is free from not only ghost instability, gradient instability and superluminality but also  Belinsky-Khalatnikov-Lifshitz (BKL) instability \cite{Belinsky:1970ew},
	\item  its predictions on the scalar spectral index and tensor-to-scalar ratio are compatible with observations \cite{Planck:2019kim}.
\end{itemize}
As we will briefly review in Section \ref{sec:model}, this model comprises of two scalar fields, one of which is responsible for bouncing dynamics and described by the exceptional subclass of DHOST Ia theories with conventional asymptotics in the far past and far future. This scalar field produces blue-tilted scalar perturbations during ekpyrotic contracting phase \cite{Khoury:2001wf,Buchbinder:2007ad,Levy:2015awa,Cook:2020oaj,Ijjas:2020dws,Ijjas:2021gkf,Ijjas:2024xxx} needed to avoid the BKL instability. It calls for the other scalar field which can generate nearly scale-invariant scalar perturbations that is converted to the curvature perturbations compatible with observations \cite{Planck:2019kim} through the entropic mechanism. The primordial tensor perturbations are highly suppressed during the ekpyrotic contracting phase so that the tensor-to-scalar ratio is negligible and naturally agrees with the observations. 

The goal of the present work is to extend the viability of the two-field DHOST bouncing model constructed in \cite{An:2025xxx} to the \textit{non-linear} level. 	
To this end, we refine the model in such a way that
\begin{itemize}
	\item its viability at the linear regime is not altered,
	\item it predicts the non-Gaussianity of the primordial curvature perturbations compatible with the observations,
	\item it does not suffer from the strong-coupling issue \cite{Leblond:2008gg,Baumann:2011dt,Koehn:2015vvy,Dehghani:2025xxx,deRham:2017ewd}.
\end{itemize}

Observational consistency of any bouncing model at the non-linear level can be investigated by using non-Gaussianity.
The main source of non-Gaussianity for the model \cite{An:2025xxx} is the non-linearity of the conversion process \cite{Fertig:2016czu,Lehners:2009ja,Lehners:2010fy,Fertig:2015ola}. And since the modes of observational interest are super-horizon during the conversion phase, only non-Gaussianity of the local type is the relevant. If all the higher-order operators are suppressed during the conversion phase, the local non-Gaussianity is well under control, so that it can be made to agree with the observations by adjusting the model parameters \cite{Fertig:2016czu}. However, for the model \cite{An:2025xxx} higher-order operators can be relevant for the conversion process, which can amplify the non-Gaussianity such that the local non-Guassianity parameter $f_{NL}$ may not satisfy  the observational constraint $ -6\leq f_{NL}\leq 4.2 $ \cite{Planck:2019kim}. This is one reason why we need to improve the model \cite{An:2025xxx}.

Another reason comes from the strong coupling issue that has to be avoided for the viability of a bouncing model at the non-linear level. Once the strong coupling is present, higher order terms in perturbation expansion of the Lagrangian are not suppressed,  which may render classical description of the bouncing dynamics inconsistent and even induce non-unitarity \cite{deRham:2017ewd}.
The model of \cite{An:2025xxx} has conventional asymptotics in the far past and far future so that it is free from the strong-coupling issue or high non-Gaussianities in the asymptotic regions, c.f. \cite{Ageeva:2018lko,Ageeva:2020buc,Ageeva:2021yik,Ageeva:2022fyq,Ageeva:2024xxx}. The strong-coupling issue can, however, arise during the bounce phase since the sound speed squared of the model becomes much smaller than $ 1 $. In many cases the ratio of the higher-order terms in perturbation expansion of the Lagrangian to the second-order terms is inversely proportional to the sound speed squared \cite{Baumann:2011dt}, and thus smallness of the sound speed signals the strong-coupling of the model. In the present work we show that the strong-coupling scale for our improved bouncing scenario is well above the characteristic background energy scale all the time to confirm that the strong-coupling issue does not arise.

The rest of this paper is organized as follows. In Section \ref{sec:model} we briefly review the original two-field DHOST bouncing scenario constructed in \cite{An:2025xxx} to refine such that all the higher-order operators are relevant only during the bounce phase and determine the allowed parameter space where the local non-Gaussianity agrees with the observations. In Section \ref{sec:strong-coupling} we evaluate the strong-coupling scale of the model and show that the model does not suffer from the strong-coupling issue. We summarize the result and conclude in Section \ref{sec:conclusion}. Following \cite{Ijjas:2016vtq}, we adopt the signature of the metric to be $ (-,+,+,+) $ and work in reduced Planck units ($ c=\hbar=1 $ and $ M_{pl}\equiv 1/\sqrt{8\pi G_N}= 1 $) throughout this paper.

\section{Two-field DHOST bounce and non-Gaussianity}\label{sec:model}

In this section, we scrutinize the model  originally proposed in \cite{An:2025xxx} closely and refine it in order to make the non-Gaussianity under control.

\subsection{General setup}

The model studied in \cite{An:2025xxx} is described by an exceptional subclass of quadratic DHOST Ia theory coupled to an extra luminal scalar field $ \chi $. Its action is given by \cite{An:2025xxx,Mironov:2020pqh,Mironov:2024xxx}
\begin{align}\label{eq:full-action}
\cS_{\text{full}}=\cS_{\phi}+\cS_{\chi}.
\end{align}
$ \cS_{\phi} $ is the action of the exceptional subclass of DHOST Ia theories, namely \cite{Langlois:2015cwa,BenAchour:2016cay,Langlois:2018dxi,Mironov:2020pqh,Mironov:2024xxx}
\begin{align}\label{eq:DHOST-action}
\cS_{\phi}=\int d^4x\sqrt{-g}\left(F(\phi,X)+ F_2(\phi,X)R[g]+\sum_{I=1}^{5}A_I(\phi,X)L_I\right),
\end{align}
where $ \phi $ is the DHOST scalar field, $ R[g] $ denotes the Ricci scalar associated with the metric $ g_{\mu\nu} $. The second part $ \cS_{\chi} $ consists of a non-minimal coupling of a luminal scalar $ \chi $ to $ \phi $ and an interaction potential between $ \phi $ and $ \chi $, namely \cite{An:2025xxx}
\begin{align}\label{eq:extra-action}
\cS_{\chi}=\int d^4x\sqrt{-g}\left( -Q(\phi)D^\mu \chi D_\mu\chi-W(\phi,\chi)\right),
\end{align}
where $ D_\mu $ stands for the covariant derivative associated with the metric $ g_{\mu\nu} $. The five elementary Lagrangians $ L_I $ $ (I=1,2,3,4,5) $ of $ \cS_{\text{DHOST}} $ are given by \cite{Langlois:2018dxi}
\begin{align*}
& X\equiv g^{\mu\nu}D_\mu\phi D_{\nu}\phi,\\
& L_1\equiv(D_\mu D_\nu \phi)(D^\mu D^\nu\phi),\quad L_2\equiv(D_\mu D^\mu\phi)^2,\quad L_3\equiv D^\mu\phi (D_\mu D_\nu\phi) D^\nu\phi (D^\rho D_\rho \phi),\\
& L_4 \equiv D^\mu \phi(D_\mu D_\nu \phi)(D^\nu D^\rho \phi)D_\rho\phi,\quad L_5\equiv [D^\mu\phi (D_\mu D_\nu\phi) D^\nu\phi]^2.
\end{align*}
The Lagrangian functions $ F $, $ F_2 $ and $ A_1 $ of $ \cS_{\phi} $ and $ Q $ and $ W $ of $ \cS_{\chi} $ are independent, while the other four Lagrangian functions $ A_I $ $ (I=2,3,4,5) $ of $ \cS_{\phi} $ are given by \cite{Langlois:2018dxi,Mironov:2020pqh,Mironov:2024xxx,An:2025xxx}
\begin{subequations}
	\begin{align}
	A_2=&\;-A_1,\label{eq:DHOST-I}\\
	A_4=&\;\frac{1}{8(F_2-XA_1)^2}\left[-16X A_1^3+4(3F_2+16XF_{2X}A_1^2)-X^2 F_2 A_3^2 \right.\nonumber\\
	&\hskip3em -(16X^2 F_{2X}-12XF_2)A_3 A_1-16F_{2X}(3F_2+4XF_{2X})A_1\nonumber\\
	&\hskip3em \left.+8F_2(XF_{2X}-F_2)A_3+48F_2F_{2X}^2\right],\label{eq:DHOST-Ia1}\\
	A_5=&\;\frac{(4F_{2X}-2A_1+X A_3)(-2A_1^2-3X A_1 A_3+4F_{2X}A_1+4F_2 A_3)}{8(F_2-XA_1)^2},\label{eq:DHOST-Ia2}\\
	A_3=&\; \frac{2 \left(A_1-2 F_{2X}\right) \left(A_1 X-2 F_2\right)}{X \left(3 A_1 X-4 F_2\right)},\label{eq:DHOST-Ia-exceptional}
	\end{align}
\end{subequations}
where the subscript $ X $ stands for the derivative with respect to $ X $. We remind that the relations \eqref{eq:DHOST-I}, \eqref{eq:DHOST-Ia1} and \eqref{eq:DHOST-Ia2} define DHOST class Ia. The exceptional subclass of DHOST Ia is further constrained by the relation \eqref{eq:DHOST-Ia-exceptional}, without which the coupling of $ \phi $ to the luminal scalar field $ \chi $ must induce an emergent superluminality \cite{Mironov:2020pqh,Mironov:2024xxx}.

We are interested on a cosmological bouncing scenario whose background metric is given by the spatially-flat Friedmann-Lema\^itre-Robertson-Walker (FLRW) metric
\begin{align}
ds^2=-dt^2+a(t)^2\delta_{ij}dx^i dx^j,\quad i,j=1,2,3,
\end{align}
where $ a(t) $ is the scale factor. In order to investigate the instability and superluminality issues we first need to perform a perturbation expansion of the action up to the second-order around the cosmological background. Since the dependence of the DHOST action $ \cS_{\text{DHOST}} $ on $ \phi $ is complicated, it is convenient to work with the so-called unitary gauge where the perturbation of the DHOST field $ \phi $ is set to be vanishing. To describe the perturbed metric, we employ the standard Arnowitt-Deser-Misner (ADM) decomposition, namely
\begin{align}\label{eq:metric-ADM}
ds^2 = -N^2 dt^2+\gamma_{ij}(dx^i+N^i dt)(dx^j+N^j dt),
\end{align}
where $ N $ and $ N^i $ are lapse and shift functions, respectively. And $\gamma_{ij}$ is the induced metric on the 3-manifold. Focusing on the scalar and tensor perturbations, the metric components can be written as
\begin{align}\label{eq:metric-perturbation}
N=1+\alpha,\quad N_i=\partial_i\beta,\quad 
\gamma_{ij}=a^2 e^{2\psi}\left(\delta_{ij}+h_{ij}^{TT}+\frac12 h_{ik}^{TT}h_j^{k}{}^{TT}\right),
\end{align}
where $ h_{ij}^{TT} $ denotes the traceless and transverse tensor perturbation. Note that in the unitary gauge the anisotropic terms in the induced metric $ \gamma_{ij} $ are gauged away. The scalar field $ \chi $ is decomposed as
\begin{align}\label{eq:chi-perturbation}
\chi(t,x)= \bar\chi(t)+\delta\chi(t,x).
\end{align}

The second-order perturbed action is then given by \cite{Mironov:2020pqh,Mironov:2024xxx,An:2025xxx} (see also Appendix \ref{app:perturbation-expansion})
\begin{align}
\cS_{\text{full}}^{(2)}=\;&\int d^4x\;a^3\left(\frac{\mathcal G_{\mathcal T}}{8}(\dot h_{ij}^{TT})^2-\frac{\mathcal F_{\mathcal T}}{8a^2}(\partial_k h_{ij}^{TT})^2+\right.\nonumber\\
&\hskip3em\left.+G_{AB}\dot w^A\dot w^B-\frac{1}{a^2}F_{AB}\delta^{ij}\partial_i w^A\partial_j w^B+\cdots\right),\label{eq:second-order-full-action}
\end{align}
where $ A,B=1,2 $, $ w^1=\tilde\psi\equiv\psi-\Delta\alpha $,  $ w^2=\delta\chi $ and the dot notation refers to the derivative with respect to $ t $. The 2-by-2 matrices $ G_{AB} $ and $ F_{AB} $ are given by
\begin{align}\label{eq:GAB-FAB}
G_{AB}=\begin{pmatrix}
\GS+\frac{\GT^2}{\tilde\Theta^2}\dot{\bar\chi}^2 Q & -\frac{\Lambda}{\tilde\Theta}\dot{\bar\chi} Q\\
-\frac{\Lambda}{\tilde\Theta}\dot{\bar\chi} Q & Q
\end{pmatrix},\quad
F_{AB}=\begin{pmatrix}
\FS & -\frac{\Lambda}{\tilde\Theta}\dot{\bar\chi} Q\\
-\frac{\Lambda}{\tilde\Theta}\dot{\bar\chi} Q & Q
\end{pmatrix},
\end{align}
where
\begin{subequations}
	\begin{align}
	\GT=\;&\cM,\\
	\FT=\;&2F_2,\\
	\GS=\;&3\GT+\frac{\GT^2\tilde\Sigma}{\tilde\Theta^2},\\
	\FS=\;&\frac{1}{a}\frac{d}{dt}\left(\frac{a\GT\Lambda}{\tilde\Theta}\right)-\FT.
	\end{align}
\end{subequations}
The explicit expressions for $ \cM $, $ \tilde\Sigma $, $ \tilde\Theta $, $ \Lambda $ and $ \Delta $ are given in the Appendix, see \eqref{eq:definition-cM-cB}, \eqref{eq:Delta0-1} and \eqref{eq:functions-general}. The ellipses in the above perturbed action \eqref{eq:second-order-full-action} denote the terms with less than two derivatives, which are irrelevant in discussing instability and superluminality issues.

The necessary and sufficient condition for the scalar sector to be free of ghost instability, gradient instability and superliminality is that the matrices $ G_{AB} $ and $ F_{AB} $ are both positive definite and all the eigenvalues of $ G^{-1}F $ are not greater than unity \cite{Mironov:2020pqh,Mironov:2024xxx,An:2025xxx}. Here the positive definiteness of the matrices $ G_{AB} $ and $ F_{AB} $ implies that
\begin{align*}
& Q>0,\\
& \det G_{AB}=\left(\GS+\frac{\GT^2}{\tilde\Theta^2}\dot{\bar\chi}^2 Q\right)Q-\left(\frac{\Lambda}{\tilde\Theta}\right)^2\dot{\bar\chi}^2 Q^2>0,\\
& \det F_{AB}=\FS Q-\left(\frac{\Lambda}{\tilde\Theta}\right)^2\dot{\bar\chi}^2 Q^2>0.
\end{align*}
Meanwhile, for the exceptional subclass of DHSOT Ia theories where the matrices $ G_{AB} $ and $ F_{AB} $ given by \eqref{eq:GAB-FAB}, the eigenvalues of $ G^{-1}F $ are simply given by
\begin{align}
c_{s-}^2\equiv 1,\quad c_{s+}^2\equiv \det F_{AB}/\det G_{AB}.
\end{align}
Combining all of these conditions, we find that the necessary and sufficient condition is
\begin{align}\label{eq:stability-nonsuper-condition}
Q>0,\quad \GS+\frac{\GT^2}{\tilde\Theta^2}\dot{\bar\chi}^2 Q\geq \FS> \left(\frac{\Lambda}{\tilde\Theta}\right)^2\dot{\bar\chi}^2 Q.
\end{align}
Similarly, the necessary and sufficient condition for the tensor sector is
\begin{align}\label{eq:stability-nonsuper-condition-tensor}
\GT\geq\FT>0.
\end{align}

At this stage, it is worth to emphasize that the above condition \eqref{eq:stability-nonsuper-condition} is less restrictive than the sufficient one given in \cite{An:2025xxx}, namely 
\begin{align}\label{eq:sufficient-condition}
Q>0,\quad \GS\geq\FS> \left(\frac{\Lambda}{\tilde\Theta}\right)^2\dot{\bar\chi}^2 Q.
\end{align}
The sufficient condition \eqref{eq:sufficient-condition} does not cause any trouble building a non-pathological bounce model compatible with observations at the linear level, as shown in \cite{An:2025xxx}. However, this is not the case for the non-linear level.   
In fact, the sufficient condition \eqref{eq:sufficient-condition} has a drawback that for the rolling scalar field $ \chi $ it does not allow the higher-order terms for $ \phi $ to be neglected, since otherwise it is violated, namely (see e.g. \eqref{eq:Lambda-Theta-Sigma-model}, neglecting the functions $ a_1 $, $ g_1 $ and $ f_2 $)
\begin{align}
\left.
\begin{aligned}
& \GT=1,\quad \tilde\Theta=H,\\
& \GS=\frac{X F_X}{H^2},\quad \FS=-\frac{\dot H}{H^2},\\
& XF_X+Q\dot{\bar\chi}^2=-\dot H,
\end{aligned}\right\}\quad\implies \quad \GS<\GS+\frac{\GT^2}{\tilde\Theta^2}\dot{\bar\chi}^2 Q= \FS,
\end{align}
where $ H\equiv \dot a/a $ is the Hubble parameter. Notice that the conversion of $ \delta\chi $ into the curvature fluctuations is activated only when the scalar field $ \chi $ rolls. 
Given that non-suppression of the higher-order terms typically leads to large non-Gaussianities disfavored by observations, it is manifest that the sufficient condition \eqref{eq:sufficient-condition} may render the model incompatible with observations at the non-linear level. 
 This indicates that the model proposed in \cite{An:2025xxx} should be refined by using the new condition \eqref{eq:stability-nonsuper-condition} in place of \eqref{eq:sufficient-condition}.

\subsection{Explicit model}

In the following, we construct an explicit bouncing model satisfying the conditions \eqref{eq:stability-nonsuper-condition} and \eqref{eq:stability-nonsuper-condition-tensor} and approaching the conventional General Relativity in the far past and far future by employing the so-called ``inverse method'' used in \cite{Ijjas:2016tpn,Ijjas:2016vtq,Libanov:2016kfc,Kolevatov:2017voe,Mironov:2018oec,Cai:2017dyi,An:2025xxx}. That is, we first specify the background evolution and then construct the independent Lagrangian functions such that the background equations of motion and the constraints are satisfied.

\begin{figure}
	\centering
	\includegraphics[width=0.7\linewidth]{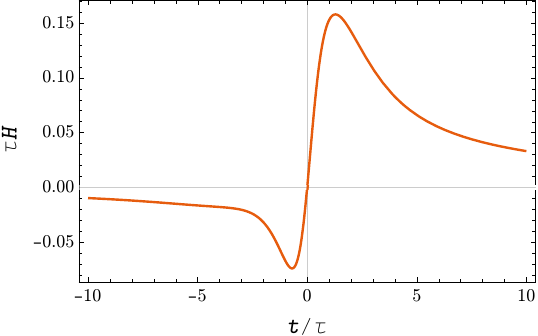}
	\caption{Plot of the Hubble parameter for $ \epsilon=10 $.}
	\label{fig:hubble}
\end{figure}

As the first step of the inverse method, we set the background evolution to be \cite{An:2025xxx}
\begin{subequations}\label{eq:bg-solution}
\begin{align}
& a(t)=\sigma \left(\frac{t}{\tau}\right)\left(\left(\frac{t}{\tau}\right)^2+1\right)^{\frac 16}+ \sigma\left(-\frac{t}{\tau}\right) \left(\left(\frac{t}{\tau}\right)^2+1\right)^{\frac{1}{2 \epsilon }},\\
& \phi(t)=t,\label{eq:bg-solution-phi}\\
& \bar\chi( t)=\bar\chi_0+\lambda \cosh^{-1}\left(\frac{ t}{\tau p}-c\right),
\end{align}
\end{subequations}
where $ \sigma(x)\equiv1/\left[1+\exp(-x)\right] $ stands for the logistic sigmoid function. The form of the scale factor $ a $ was chosen such that the bounce occurs at $ t=0 $ and the asymptotic behavior of the Hubble parameter $ H $ becomes (see Fig. \ref{fig:hubble})
\begin{subequations}\label{eq:H-asymptotics}
	\begin{align}
	& H\to \frac{1}{3t}\quad \text{for}\quad t\to+\infty,\\
	& H\to \frac{1}{\epsilon t}\quad \text{for}\quad t\to-\infty.
	\end{align}
\end{subequations}
The parameter $ \epsilon $ controls the equation of state during the ekpyrotic contraction phase and should be set to be greater than $ 3 $ in order to avoid the BKL instability during the contracting phase \cite{Khoury:2001wf,Buchbinder:2007ad,Levy:2015awa}. The parameter $ \tau\gg 1 $ specifies the duration of the bounce phase (i.e. $ -\tau\lesssim t\lesssim \tau $) and is related to the amplitude of the power spectra for the scalar perturbations.

The form of the function $ \bar\chi $ was chosen such that the scalar field $ \chi $ rolls only for a certain period around $ t=\tau cp $ during which the entropic perturbations $ \delta\chi $ are converted into the curvature fluctuations. It is desirable for the conversion phase not to overlap with the bounce phase due to their complexity and non-linearity. Furthermore, it is beneficial to make the conversion occur \emph{after} the bounce phase particularly from the perspective of non-Gaussianity \cite{Fertig:2016czu}. We will choose the parameters $ c $ and $ p $ such that the conversion phase comes long after the bounce phase. The parameter $ \lambda $ specifies the average velocity of $ \chi $ during the conversion phase. The bias $ \bar\chi_0>0 $ was introduced to make $ \bar\chi $ always positive, which is necessary in order to make the interaction between $ \phi $ and $ \chi $ non-singular, as we will see soon. 

The second step of the inverse method is to determine the independent Lagrangian functions $ F $, $ F_2 $, $ A_1 $, $ Q $ and $ W $ of \eqref{eq:full-action} such that the background equations of motion \eqref{eq:bg-eqns} and the constraints \eqref{eq:stability-nonsuper-condition} and \eqref{eq:stability-nonsuper-condition-tensor} are satisfied. We choose the ansatz for these independent Lagrangian functions as 
\begin{subequations}
	\begin{align}
	& F(\phi,X)=f_0\left(\phi\right)- X f_1\left(\phi\right)+  X^2f_2\left( \phi\right),\\
	& F_2(\phi,X)=\frac12-g_1\left(\phi\right)- Xg_1\left(\phi\right),\\
	& A_1(\phi,X)=-a_1\left(\phi\right)- X a_1\left(\phi\right),\\
	& Q(\phi) = \frac12\frac{\sigma \left(-\frac{\phi}{\tau}\right)}{ \left|\frac{\phi}{\tau}\right|^{2(1+b)}+1}+\frac{1}{2}\sigma \left(\frac{\phi}{\tau}\right),\\
	& W(\phi,\chi)=W_2(\phi)\chi^2.
	\end{align}\label{eq:ansatz}
\end{subequations}
The form of $ F_2 $ and $ A_1 $ was chosen to satisfy \cite{An:2025xxx,Mironov:2018oec}
\begin{align}
\GT=\FT=1,
\end{align}
so that the tensor sector is free of ghost and gradient instabilities and the tensor modes propagate at the speed of unity. 

The function $ Q(\phi) $ was chosen such that during the ekpyrotic contracting phase power spectrum of the perturbations $ \delta\chi $ becomes nearly scale-invariant with the spectral index \cite{An:2025xxx,Qiu:2013eoa,Li:2013hga,Fertig:2013yyy,Fertig:2016czu}
\begin{align}\label{eq:ns}
n_s=1-(2b\epsilon)/(\epsilon-1),
\end{align}
and after the bounce phase the kinetic term of $ \chi $ becomes canonical. Note that this way of generating the nearly scale-invariant scalar perturbations has a merit; it produces negligible intrinsic non-Gaussianities \cite{Fertig:2013yyy,Fertig:2015ola,Fertig:2016czu}. Due to the nature of the entropic mechanism, this spectral index $ n_s $ will be inherited to the density perturbations, and hence it is subjected to the observational constraint \cite{Planck:2019kim}
\begin{align}
0.95\lesssim n_s\lesssim 0.97.
\end{align}
Notice that for any value of $ \epsilon $ one can adjust the parameter $ b $ in order for the above constraint to be satisfied.

Only three out of the six model functions ($ f_{0,1,2} $, $ a_1 $, $ g_1 $ and $ W_2 $) in the ansatz \eqref{eq:ansatz} are independent due to the background equations of motion \eqref{eq:bg-eqns}. Inserting the ansatz \eqref{eq:ansatz} into the background equations of motion \eqref{eq:bg-eqns} and taking into account \eqref{eq:bg-solution-phi}, we obtain
\begin{subequations}\label{eq:bg-eqns-model}
	\begin{align}
	0=\;& f_0(t)+f_1(t)+f_2(t)+3 H^2+2 \dot H+Q\dot{\bar\chi}^2-W_2(t)\bar\chi^2 ,\label{eq:bg-eqn-psi}\\
	0=\;& f_0(t)-f_1(t)-3 f_2(t)+3H^2 \left(4 a_1(t)-2 g_1(t)+1\right)-6g_1(t) \dot H\nonumber\\
	&\hskip5em+6 H \dot g_1-Q\dot{\bar\chi}^2- W_2(t)\bar\chi^2,\label{eq:bg-eqn-alpha}\\
	0=\;& Q(t)\ddot{\bar\chi}+\left(3HQ(t)+\dot Q(t)\right)\dot{\bar\chi}+W_2(t)\bar\chi.\label{eq:bg-eqn-chi}
	\end{align}
\end{subequations}
It follows from the background equation of motion \eqref{eq:bg-eqn-chi} that the model function $ W_2(\phi) $ is completely determined by the background functions \eqref{eq:bg-solution}. Since $ \bar\chi $ is always positive, the model function $ W_2(t) $ never diverges.

Since the model functions $ f_2 $, $ a_1 $ and $ g_1 $ characterize the higher-order terms of the action, it is convenient to make them independent. Indeed, we will choose these functions such that they are relevant only during the bounce phase and exponentially suppressed in the far past and far future.  The remaining model functions $ f_0 $ and $ f_1 $ are fully determined in terms of $ f_2 $, $ a_1 $ and $ g_1 $; it follows from \eqref{eq:bg-eqn-psi} and \eqref{eq:bg-eqn-alpha} that
\begin{subequations}
\begin{align}\label{eq:f0f1}
& f_0=-\dot H-3H^2+f_2-3 H^2 \left(2 a_1-g_1\right)+3g_1 \dot{H}-3 H \dot{g}_1+W_2\bar\chi^2,\\
& f_1=-\dot H-2f_2+3H^2(2a_1-g_1)-3g_1\dot H+3 H \dot g_1-Q\dot{\bar\chi}^2.\label{eq:f1}
\end{align}
\end{subequations}

Now the remaining task is to choose the model functions $ f_2 $, $ a_1 $ and $ g_1 $ such that the necessary and sufficient condition \eqref{eq:stability-nonsuper-condition} for the scalar sector to be free from ghost instability, gradient instability and superluminality is satisfied. Substituting the ansatz \eqref{eq:ansatz} into \eqref{eq:functions-general} and taking into account \eqref{eq:f1}, we obtain
\begin{subequations}\label{eq:Lambda-Theta-Sigma-model}
	\begin{align}
	\Lambda=\;&1-3g_1,\\
	\tilde\Theta=\;& H(1+4a_1+g_1)+\dot g_1,\\
	\tilde\Sigma=\;&-\dot H-3H^2+4 f_2-Q\dot{\bar\chi }^2-6 g_1 \ddot{g}_1+3 \dot{g}_1{}^2 \nonumber\\
	&-3 H^2 \left(9 a_1 g_1+13 a_1+3 g_1^2+2 g_1\right)+\dot{H} \left(-9 a_1 g_1+3 a_1-3 g_1^2+6 g_1\right)\nonumber\\
	&-3 H \left(-5 a_1 \dot{g}_1+3 g_1 \dot{a}_1+6 g_1 \dot{g}_1-\dot{a}_1+4 \dot{g}_1\right).\label{eq:tildeSigma-Model}
	\end{align}
\end{subequations}
Inserting the above expressions into the condition \eqref{eq:stability-nonsuper-condition} gives
\begin{subequations}
\begin{align}
&  \frac{d}{dt}\left(\frac{1-3g_1}{H(1+4a_1+g_1)+\dot g_1} \right)-\frac{4H(a_1+g_1)+\dot g_1}{H(1+4a_1+g_1)+\dot g_1}\nonumber\\
&\hskip10em>Q(t)\dot{\bar\chi}^2 \left(\frac{1-3g_1}{H(1+4a_1+g_1)+\dot g_1}\right)^2,\label{eq:condition-a1-g1}\\
& f_2\geq \frac{1}{4} \dot H \left(21 a_1 g_1-7 a_1+6 g_1^2-4 g_1\right)+\frac{1}{4} \left(9 g_1 \ddot g_1-\ddot g_1-10 \dot g_1^2\right)\nonumber\\
&\hskip3em+\frac{1}{4} H \left(-59 a_1 \dot g_1+21 g_1 \dot a_1+7 g_1 \dot g_1-7 \dot a_1+\dot g_1\right)\nonumber\\
&\hskip3em +\frac{1}{4} H^2 \left(-17 a_1 g_1-64 a_1^2+11 a_1+2 g_1^2-4 g_1\right).\label{eq:f2-lower-bound}
\end{align}
\end{subequations}

Let us first consider the phase, during which the extra scalar $ \bar\chi $ does not roll. In this case, the condition \eqref{eq:condition-a1-g1} can be written as
\begin{align}
\frac{d}{dt}\left(\frac{1-3g_1}{H(1+4a_1+g_1)+\dot g_1} \right)-\frac{4H(a_1+g_1)+\dot g_1}{H(1+4a_1+g_1)+\dot g_1}>0,\label{eq:condition-a1-g1-simple}
\end{align}
The condition \eqref{eq:condition-a1-g1-simple} is the same as in \cite{An:2025xxx}, so we make the same choice for the model functions $ a_1 $ and $ g_1 $ as in \cite{An:2025xxx}, namely
\begin{subequations}\label{eq:g1-a1}
	\begin{align}
	& g_1(t)=\frac{w}{3} \cosh^{-2}\left(\frac{t}{\tau}+u\right),\\
	& a_1(t)=\frac{w}{12}\cosh^{-2}\left(\frac{t}{\tau}+u\right) \left(2\frac{\left(\frac{t}{\tau}\right)^2 \tanh \left(\frac{t}{\tau}+u\right)+\tanh \frac{t}{\tau}}{  \left(\left(\frac{t}{\tau}\right)^2+1\right)\tau H}-1\right),
	\end{align}
\end{subequations}
where the parameters $ u>0 $ and $ w>1 $ were introduced to avoid the fine-tuning issue \cite{Mironov:2018oec}.

To choose the form of the model function $ f_2 $, we first observe the asymptotic behavior of the right-hand side of the condition \eqref{eq:f2-lower-bound}, namely
\begin{align}
\text{RHS of \eqref{eq:f2-lower-bound}}\to \frac{w}{4\tau^2}\cosh^{-2}\left(\frac{t}{\tau}+u\right) \text{ as } |t|\to+\infty.
\end{align}
We therefore select the model function $ f_2(t) $ as 2$\tau^2$
\begin{align}\label{eq:f2}
f_2(t)=\frac{w}{2\tau^2}\cosh^{-2}\left(\frac{t}{\tau}+u\right),
\end{align}
which is exponentially suppressed in the far past and far future.

\begin{figure}
	\centering
	\subfigure{\includegraphics[width=0.45\linewidth]{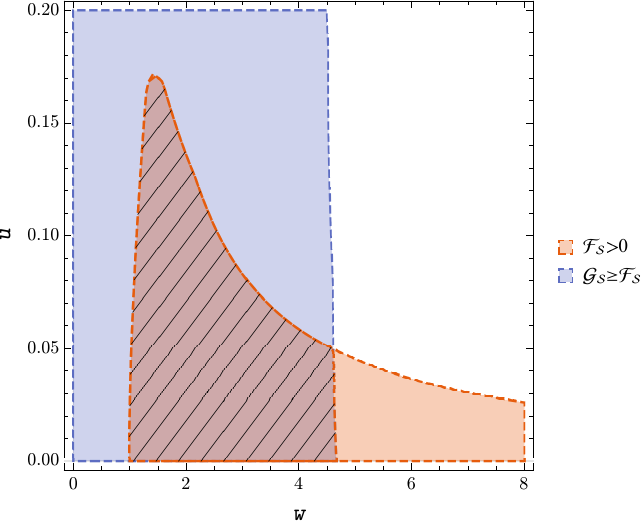}}
	\subfigure{\includegraphics[width=0.45\linewidth]{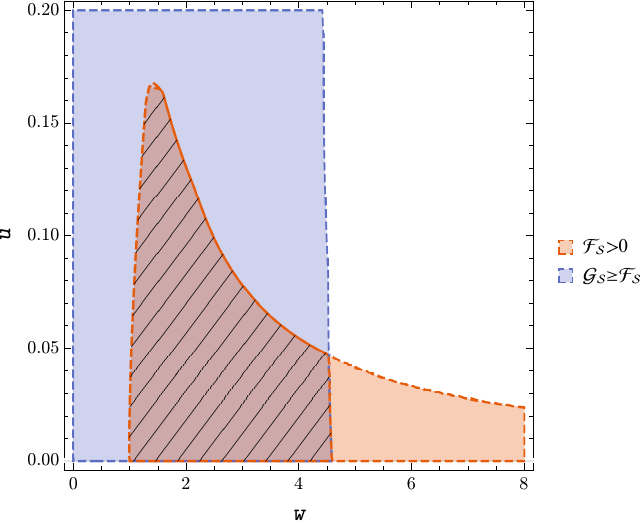}}
	\subfigure{\includegraphics[width=0.45\linewidth]{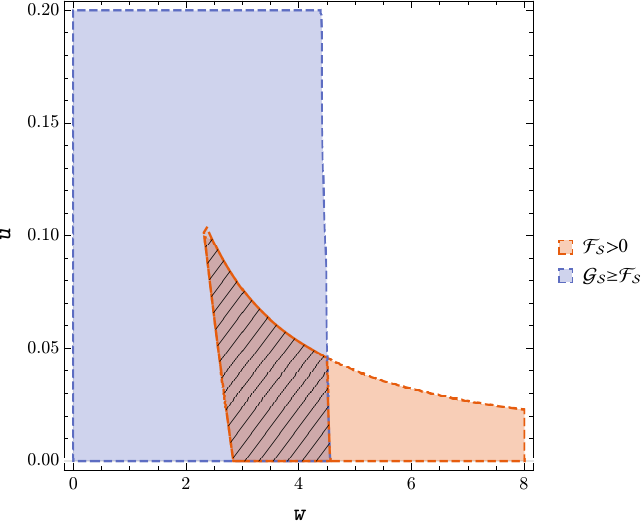}}
	\subfigure{\includegraphics[width=0.45\linewidth]{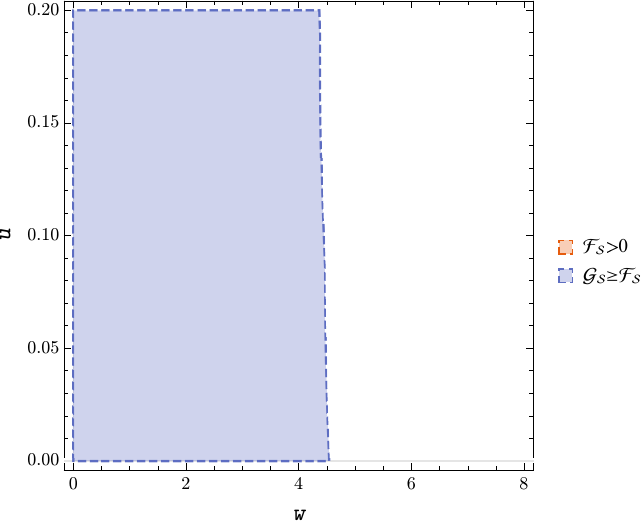}}
	\caption{Allowed region of $w$ and $u$ for $\epsilon=10$, $ \epsilon=12 $, $ \epsilon=13 $ and $ \epsilon=14 $.}
	\label{fig:w-u-allowed}
\end{figure}

Note that the model functions \eqref{eq:g1-a1} and \eqref{eq:f2} were chosen by merely matching the asymptotic behaviours, so that they do not guarantee satisfaction of the conditions \eqref{eq:condition-a1-g1-simple} and \eqref{eq:f2-lower-bound} at all times. We have numerically determined the allowed region in the parameter space of $ w $ and $ u $ for various values of $ \epsilon $, see the intersection regions filled with slash black lines in Fig. \ref{fig:w-u-allowed}.  One can readily see that the allowed region shrinks as $ \epsilon $ gets greater and finally becomes empty for $ \epsilon\geq 14 $.

\begin{figure}
	\centering
	\subfigure[Log-plot of $\tilde\Theta^2(\GS-\FS)$]{\includegraphics[width=0.45\linewidth]{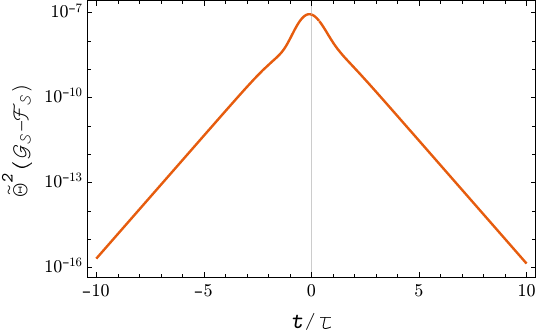}}
	\subfigure[Log-plot of $\tilde\Theta^2\GS$ and $\tilde\Theta^2\FS$]{\includegraphics[width=0.5\linewidth]{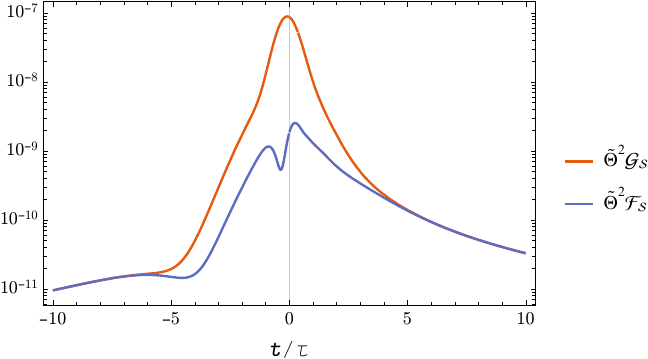}}
	\caption{Log-plot of $\tilde\Theta^2(\GS-\FS)$, $\tilde\Theta^2 \GS$ and $\tilde\Theta^2\FS$ for the parameters $ \epsilon $, $ w $ and $ u $ given by \eqref{eq:parameters-epsilon-w-u}. These two plots confirm $ \GS\geq\FS>0 $.}
	\label{fig:theta2gs-fs}
\end{figure}

\begin{figure}
	\centering
	\includegraphics[width=0.7\linewidth]{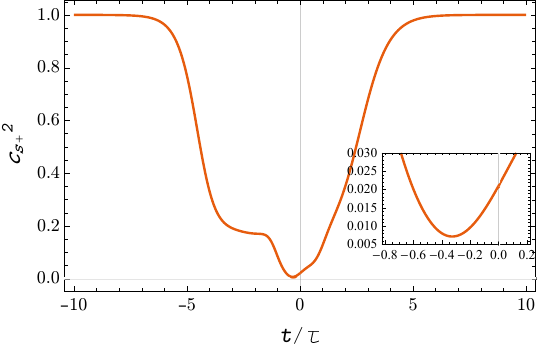}
	\caption{Plot of sound speed squared $ c_{s+}^2 $ for the parameters given by \eqref{eq:parameters-epsilon-w-u}, which shows that $ c_{s+}^2 $ is always greater than zero and less than one. The inset shows a magnification of the small region around the minimum of $ c_{s+}^2 $.}
	\label{fig:cssquared}
\end{figure}

Since the parameters $ \epsilon $, $ w $ and $ u $ are related only to the stability issue and are not connected to the observational quantities, we fix them as (c.f. \cite{Mironov:2018oec,An:2025xxx})
\begin{align}\label{eq:parameters-epsilon-w-u}
\epsilon=10,\quad w=2,\quad u=\frac{1}{10},
\end{align}
which belongs to the allowed region in Fig. \ref{fig:w-u-allowed}. We confirmed that the conditions \eqref{eq:condition-a1-g1-simple} and \eqref{eq:f2-lower-bound} are satisfied for the model functions \eqref{eq:g1-a1} and \eqref{eq:f2} with the model parameters $ \epsilon $, $ w $ and $ u $ given by \eqref{eq:parameters-epsilon-w-u}, see Fig. \ref{fig:theta2gs-fs} and also Fig. \ref{fig:cssquared} showing the sound speed squared $ c_{s+}^2=\FS/\GS $ greater than zero and less than one.

\begin{figure}
	\centering
	\includegraphics[width=0.7\linewidth]{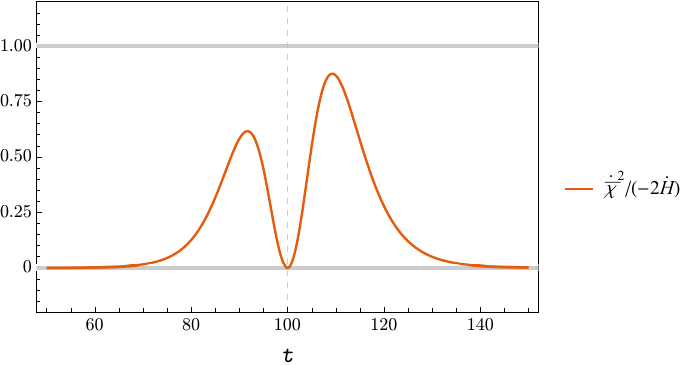}
	\caption{Plot of $\dot{\bar\chi}^2/(-2\dot H)$ for $ \lambda=0.14 $ and $ c=p=10 $.}
	\label{fig:condition-chibar-satisfying}
\end{figure}

Now we consider the conversion phase. As mentioned before, the parameters $ p $ and $ c $ will be chosen such that the conversion phase is well-separated from the bounce phase, and therefore for the rolling scalar $ \chi $ the condition \eqref{eq:condition-a1-g1} is reduced to the following inequality
\begin{align}\label{eq:condition-chibar}
\dot{\bar\chi}^2<-2\dot H,
\end{align}
which is translated into
\begin{align}
\sqrt{\frac32}\lambda \frac{t}{\tau p}\frac{\left|\tanh\left(\frac{t}{\tau p}-c\right)\right|}{\cosh\left(\frac{t}{\tau p}-c\right)}< 1.
\end{align}
The above inequality is satisfied when
\begin{align}
\lambda\leq \frac{\sqrt{\frac23}}{\frac{c}{2}+0.6}.\label{eq:condition-lambda}
\end{align}
For instance, the inequality \eqref{eq:condition-lambda} holds if $ \lambda\leq 0.14 $ for $ c=10 $, see Fig.\ref{fig:condition-chibar-satisfying}. Here the values of the parameters $ c $ and $ p $ were chosen such that the conversion  occurs long after the bounce phase.

In summary, our scenario specified by \eqref{eq:ansatz}, \eqref{eq:bg-eqn-chi}, \eqref{eq:f0f1}, \eqref{eq:f1}, \eqref{eq:g1-a1} and \eqref{eq:f2} is free from ghost instability, gradient instability and superluminality around the bouncing background \eqref{eq:bg-solution} at least for the parameters
\begin{align}\label{eq:parameters-ep-w-c-p-lambda}
\epsilon=10,\quad w=2,\quad u=\frac{1}{10}, \quad c=p=10,\quad \lambda\leq 0.14.
\end{align}

\subsection{Conversion and local non-Gaussianity}

So far, we have refined the two-field DHOST bounce scenario such that it is very close to conventional GR with canonical scalar fields except for a certain period around the bounce phase, maintaining stability and non-superluminality of its original version \cite{An:2025xxx}. Now, we would like to see if its predictions on the quantities related the primordial density perturbations can be compatible with the observations.

Since the conversion occurs long after the bounce phase for the parameters \eqref{eq:parameters-ep-w-c-p-lambda}, all the higher-order terms can be neglected during the conversion phase. In particular, because the modes of interest are super-horizon during the conversion phase, one has only to consider the large-scale evolution of the perturbations and non-Gaussianity of the local type. Therefore, results of the previous work \cite{Fertig:2016czu} on the conversion after the bounce can be straightforwardly applied to our model. However, it is instructional to check the phenomenological aspect of our improved model from the perspective of the inverse method.

\begin{figure}
	\centering
	\subfigure{\includegraphics[width=0.45\linewidth]{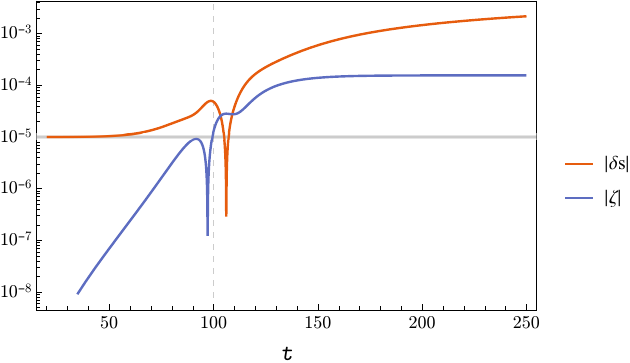}}
	\subfigure{\includegraphics[width=0.5\linewidth]{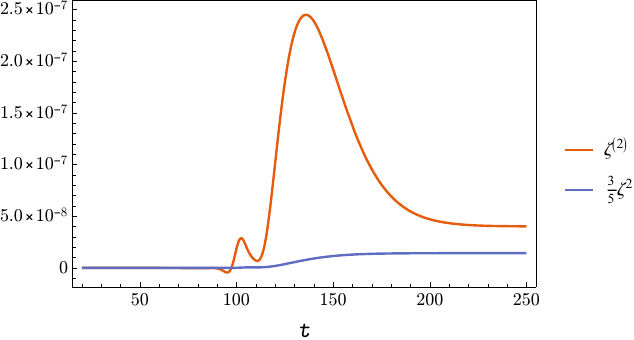}}
	\caption{Log-plot of linear perturbations $|\delta s|$ and $|\zeta|$ (left panel) and plot of second-order perturbation $ \zeta^{(2)} $ in comparison with $ \frac35\zeta^2 $ (right panel).}
	\label{fig:ls-evolution-linear-second}
\end{figure}

As mentioned in the previous subsection, the scalar spectral index can be easily made to agree with the observations by adjusting the parameter $ b $. We therefore focus on the amplitude and local non-Gaussianity. The estimation of these quantities requires solving the large-scale equations of the perturbations, which have been extensively studied employing the so-called covariant formalism in the literature \cite{Langlois:2006vv,Fertig:2015ola,Ijjas:2014fja,Fertig:2016czu}. These equations are given for the gauge-invariant curvature perturbation $ \zeta $ and entropic fluctuation $ \delta s $, which are briefly reviewed in Appendix \ref{app:LS-evolution}. We have numerically solved the evolution equations \eqref{eq:perturbation-EOM} for first-order perturbations $ \zeta $, $ \delta s $ and second-order perturbations $ \zeta^{(2)} $, $ \delta s^{(2)} $ for the fixed parameters $ c=p=10 $ and free parameters $ \lambda\leq 0.14 $ and $ \bar\chi_0>0 $. For instance, for the parameters
\begin{align}
\lambda=0.14,\quad \bar\chi_0=0.029,
\end{align}
the evolution of first-order and second-order perturbations is shown in Fig. \ref{fig:ls-evolution-linear-second}.

\begin{figure}
	\centering
	\includegraphics[width=0.7\linewidth]{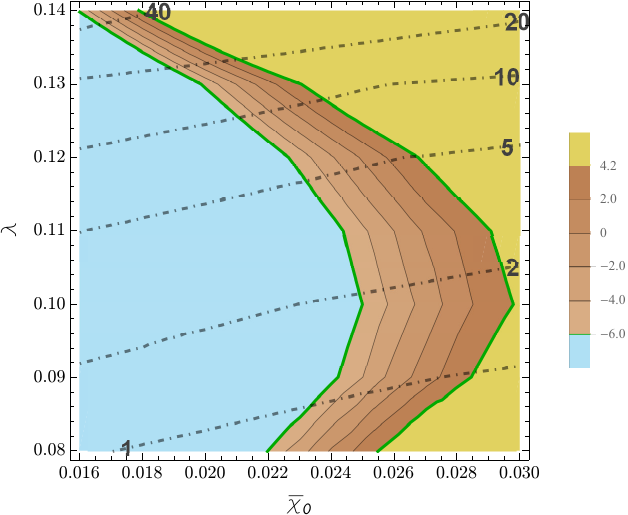}
	\caption{Contour-plot of $ f_{NL} $ and $E_{conv}$ for various values of $\bar\chi_0$ and $\lambda$. The light-blue region indicates $ f_{NL}<-6.0 $, while the dark-yellow region indicates $ f_{NL}>4.2 $. The region consistent with Planck data \cite{Planck:2019kim} is shaded with brown and its boundary is plotted with green. The conversion efficiency $ E_{conv} $ is plotted by dot-dashed lines with labels showing the value of $ E_{conv} $.}
	\label{fig:econv-fnl}
\end{figure}

After solving the evolution equations, we evaluate the conversion efficiency $ E_{conv} $ and local non-Gaussianity $ f_{NL} $, which are defined by \cite{Lehners:2009ja,Fertig:2015ola,Fertig:2016czu,An:2025xxx}
\begin{align}
E_{conv}=\left|\frac{\zeta(t_f)}{\delta s(t_i)}\right|,\quad f_{NL}=\frac53 \frac{\int_{t_i}^{t_f}\dot{\zeta}^{(2)}dt}{\left(\int_{t_i}^{t_f}\dot{\zeta} dt\right)^2},
\end{align}
with $ t_i $ and $ t_f $ respectively denoting the initial and final moment of the conversion phase. 
When ignoring $ \cO(1) $ factors, the conversion efficiency is connected to the amplitude $ A_s $ of the scalar fluctuations as \cite{An:2025xxx}
\begin{align}
A_s\simeq E_{conv}^2 \tau^{-2}.
\end{align}
Fig. \ref{fig:econv-fnl} illustrates the allowed region in the parameter space of $ \bar\chi_0 $ and $ \lambda $. The brown shaded region indicates the area where the local non-Gaussianity $ f_{NL} $ is consistent with the observations $ -6\leq f_{NL}\leq 4.2 $ \cite{Planck:2019kim}. The light blue and dark yellow shaded regions correspond to $ f_{NL}<-6 $ and $ f_{NL}>4.2 $, respectively. The conversion efficiency is plotted by dot-dashed lines with labels. One can readily see that the conversion efficiency ranges from $ \cO(1) $ to $ \cO(10) $, implying that the amplitude of the scalar power spectrum agrees with the observations $ A_s=(2.105\pm 0.030)\times 10^{-9} $ \cite{Planck:2019kim} when the parameter $ \tau $ lies in the range from $ \cO(10^{4}) $ to $ \cO(10^{5}) $.

Fig. \ref{fig:econv-fnl} shows that the conversion efficiency can be enhanced by either increasing $ \lambda $ or decreasing $ \bar\chi_0 $, and the local non-Gaussianity can agree with the observations $ -6\leq f_{NL}\leq 4.2 $ \cite{Planck:2019kim} only for a certain range of $ \bar\chi_0 $ with $ \lambda $ being fixed. One can also notice that higher conversion efficiency is not always beneficial to compatibility of the local non-Gaussianity with the observations.

\section{Strong-coupling scale}
\label{sec:strong-coupling}

In the last section, we have built an explicit bouncing scenario where its predictions on primordial density perturbations are compatible with the cosmological observations and it never exhibits ghost instability, gradient instability and superluminality. Higher-order terms are exponentially suppressed as $ |t|\to+\infty $ and away from the bounce phase the scenario is well-described by conventional scalar fields coupled to GR.

Physics during the bounce phase is, however, quite non-trivial and higher-order terms may generate considerable non-linear effects to spoil the background evolution and even lead to violation of unitarity. This worry is based on the previous studies (see e.g. \cite{Leblond:2008gg,Baumann:2011dt}) indicating that the interaction terms in the perturbation expansion of the Lagrangian are usually accompanied by a factor $ 1/c_s^2 $, while in our model the sound speed squared becomes much smaller than unity during the bounce phase, see Fig. \ref{fig:cssquared}.

\begin{figure}
	\centering
	\includegraphics[width=0.7\linewidth]{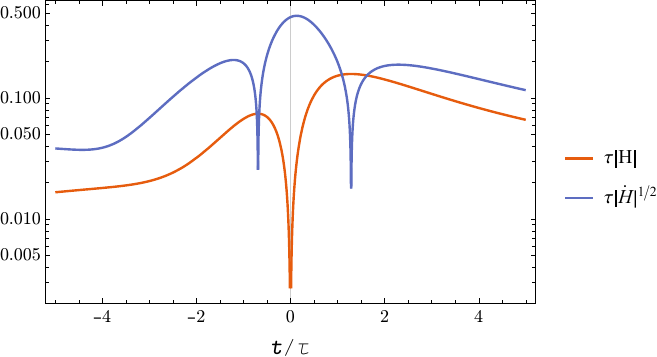}
	\caption{Log-plot of $|H|$ and $|\dot H|^{1/2}$.}
	\label{fig:eback}
\end{figure}

In general, whether the non-linear effect is strong or not is judged by comparing the characteristic background energy scale (see e.g. \cite{deRham:2017ewd,Dehghani:2025xxx})
\begin{align}
E_{\text{back}}\sim \max\left\{|H|,|\dot H|^{1/2},\cdots\right\}
\end{align}
with the strong coupling scale at which the third-order Lagrangian is comparable to the second-order one \cite{Leblond:2008gg,Baumann:2011dt,Dehghani:2025xxx}. For instance, the strong coupling scale is order of $ \sqrt{\frac{2}{\epsilon}} $ during the ekpyrotic contracting phase driven by a canonical scalar field \cite{Koehn:2015vvy}, which is much higher than $ E_{\text{back}} \sim \tau^{-1}$, provided that $ \max\left\{\tau|H|,\tau|\dot H|^{1/2},\cdots\right\}\leq \cO(1) $, $ \epsilon\lesssim \cO(10^1) $ and $ \tau\sim \cO(10^{4}- 10^5) $, see also Fig. \ref{fig:eback}.

The aim of this section is to estimate the strong-coupling scale up to the bounce phase to confirm consistency of the classical description of the background bouncing cosmology.

\subsection{Evolution of curvature perturbation}

The strong-coupling scale can be evaluated by either comparing the third-order Lagrangian with the second-order one \cite{Leblond:2008gg,Baumann:2011dt,Koehn:2015vvy} or computing the equilateral bispectrum \cite{Dehghani:2025xxx}. To this end, we begin by finding the evolution of curvature perturbation up to the bounce phase.

Since before and during the bounce phase the extra luminal scalar field $ \chi $ is effectively decoupled, we focus only on the $ \phi $ sector, where the relevant model parameters are $ \epsilon $, $ u $, $ w $ and $ \tau $. The others except for $ \tau $ have already been fixed in the previous section, see \eqref{eq:parameters-epsilon-w-u}. As the most conservative choice, we now fix the parameter $ \tau $ to be 
\begin{align}
\tau=10^{4}.
\end{align}

Since the scalar field $ \chi $ is decoupled, the constraints \eqref{eq:constraints-full} are reduced to
\begin{subequations}\label{eq:constraints-alpha-beta}
	\begin{align}
	& \alpha=\frac{\cM\dtpsi}{\tilde\Theta},\\
	& \frac{1}{a^2}\beta_{,ii}=3\dtpsi+\frac{1}{\tilde\Theta}\left(\tilde\Sigma\alpha-\Lambda\frac{\tilde\psi_{,ii}}{a^2}\right),
	\end{align}
\end{subequations}
and the unconstrained second-order action becomes
\begin{align}
\cS_{\phi}^{(2)}[\tilde\psi]=\int d^3\mathbf x dt\; a^3\left[\cM\left( 3\tilde\Theta^2+\cM\tilde\Sigma\right)\left(\frac{\dtpsi}{\tilde\Theta}\right)^2 +\frac{2 F_2 \tilde{\psi }_{,i}\tilde{\psi }_{,i}}{a^2}-2\cM\Lambda\frac{\dot{\tilde{\psi }}}{\tilde\Theta}\frac{  \tilde{\psi }_{,ii} }{a^2}\right].\label{eq:second-order-action-nonsingular}
\end{align}
Implementing integration by parts for the third term in \eqref{eq:second-order-action-nonsingular} gives 
\begin{align}
\cS_{\phi}^{(2)}[\tilde\psi]=\int d^3\mathbf x dt\; a^3 \left(\GS\dtpsi^2 -\FS\frac{ \tilde{\psi }_{,i}\tilde{\psi }_{,i}}{a^2}\right).\label{eq:second-order-action-simple}
\end{align}
Let us write $ \tilde\psi $ in Fourier space as
\begin{align}
\tilde\psi(t,\mathbf x)=\frac{1}{(2\pi)^3}\int d^3\mathbf k\;\tilde\psi_{\mathbf k}(t)e^{i\mathbf k\cdot \mathbf x},\quad \tilde\psi_{\mathbf k}(t)=u_k(t)a_{\mathbf k}+u^*_k(t)a_{-\mathbf k}^\dagger,
\end{align}
where $ a_{\mathbf k} $ and $ a_{\mathbf k}^\dagger $ are time-independent annihilation and creation operators,
\begin{align}
[a_{\mathbf k},a_{\mathbf k'}^\dagger]=(2\pi)^3 \delta^{(3)}(\mathbf k-\mathbf k'),\quad [a_{\mathbf k},a_{\mathbf k'}]=[a^\dagger_{\mathbf k},a^\dagger_{\mathbf k'}]=0.
\end{align}
The evolution equation for Fourier-mode $ u_k(t) $ is given by
\begin{align}\label{eq:EOM-tildepsi}
\ddot u_k+\left(3H+\frac{\dot\GS}{\GS}\right)\dot u_k+\frac{k^2}{a^2}c_{s+}^2 u_k=0.
\end{align}

Two comments are in order. First, although the evolution equation \eqref{eq:EOM-tildepsi} for $ \tilde\psi $ becomes singular when $ \tilde\Theta $ crosses zero, its solution $ \tilde\psi $ as well as other perturbation variables $ \alpha $ and $ \beta $ still remains regular \cite{Mironov:2018oec,Koehn:2015vvy}. Second, even though the two actions give the same equation of motion \eqref{eq:EOM-tildepsi}, the integrand of \eqref{eq:second-order-action-nonsingular} is always finite \cite{Koehn:2015vvy,Battarra:2014tga}, while the integrand of \eqref{eq:second-order-action-simple} is divergent at zero point of $ \tilde\Theta $. For this reason, we regard the non-singular action \eqref{eq:second-order-action-nonsingular} fundamental and use it to compare with the third-order action.

 Introducing the Mukhanov-Sasaki variable $ v_k $ defined by $ v_k\equiv a\sqrt{\GS}u_k $, we can rewrite eq. \eqref{eq:EOM-tildepsi} as
\begin{align}\label{eq:EOM-vk}
v_k''+\left(c_{s+}^2 k^2-\frac{(a\sqrt{\GS})''}{a\sqrt{\GS}}\right)v_k=0,
\end{align}
where $ \prime $ denotes the derivative with respect to the conformal time $ \eta $ defined by $ d\eta=dt/a $. 
The solution to \eqref{eq:EOM-tildepsi} during the ekpyrotic contracting phase has been well-known.
Since the coefficient of the second term in the above equation approaches to $ k^2 $ as $ \eta\to-\infty $, we can impose the Bunch-Davies vacuum condition, namely
\begin{align}\label{eq:BD-initial-condition}
\lim_{\eta\to-\infty}v_k=\frac{1}{\sqrt{2k}}e^{-ik\eta}.
\end{align}
Using $ -\frac{\eta}{\tau}\simeq\frac{\epsilon}{\epsilon-1}(-t/\tau)^{\frac{\epsilon-1}{\epsilon}} $ and $ \GS=\FS\simeq \epsilon $ during the ekpyrotic contracting phase, the solution to \eqref{eq:EOM-vk} is given by
\begin{align}
v_k\simeq \frac{\sqrt{-\pi\eta}}{2}H_\nu^{(1)}(-k\eta),
\end{align}
or
\begin{align}
u_k\simeq\;& \frac{1}{2a\sqrt{\GS}}\sqrt{\frac{\pi\tau\epsilon}{\epsilon-1}}\left(-\frac{t}{\tau}\right)^{\frac{\epsilon-1}{2\epsilon}} H_\nu^{(1)}\left(k\frac{\epsilon\tau}{\epsilon-1}\left(-\frac{t}{\tau}\right)^{\frac{\epsilon-1}{\epsilon}} \right)\nonumber\\
\simeq\;&\frac{1}{2}\sqrt{\frac{\pi\tau}{\epsilon-1}}\left(-\frac{t}{\tau}\right)^{\frac{\epsilon-3}{2\epsilon}} H_\nu^{(1)}\left(k\frac{\epsilon\tau}{\epsilon-1}\left(-\frac{t}{\tau}\right)^{\frac{\epsilon-1}{\epsilon}} \right),
\end{align}
where $ \nu=\frac{\epsilon-3}{2(\epsilon-1)} $ and $ H_\nu^{(1)} $ is the Hankel function of the first kind.

\begin{figure}
	\centering
	\subfigure{\includegraphics[width=0.45\linewidth]{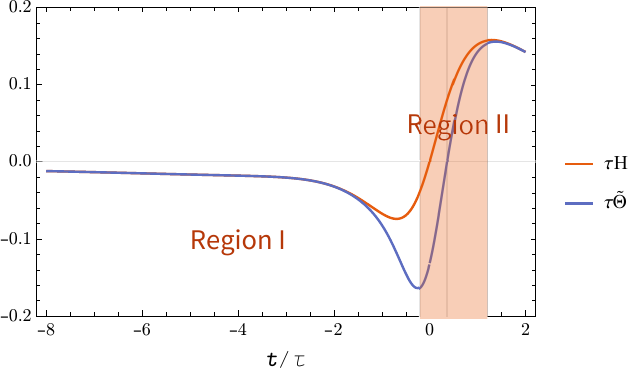}}
	\subfigure{\includegraphics[width=0.5\linewidth]{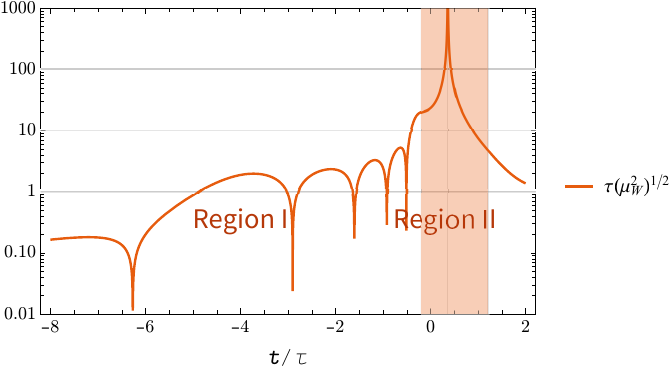}}
	\caption{Plots of Region I and II. Region II, plotted with orange background, is the interval around the zero point of $ \tilde\Theta $ where $ \tilde\Theta $ grows linearly. In Region II, the WKB approximation does not hold even in high $ k $ limit. The gray vertical lines on two panels indicate $ t_0 $.}
	\label{fig:tildethetahregions}
\end{figure}

The evolution of $ u_k $ is not simple during the bounce phase. In particular, for $ k $ such that $ k^2\sim \mu^2_W\equiv \frac{1}{c_{s+}^{2}}\left|\frac{(a\sqrt{\GS})''}{a\sqrt{\GS}}\right| $, $ u_k $  nontrivially depends on the profile of $ \GS $. But once $ k^2\gg\mu^2_W$, the solution to \eqref{eq:EOM-vk} can be obtained by the Wentzel-Kramers-Brillouin (WKB) approximation, and therefore we have
\begin{align}\label{eq:RegionI-WKB-solution}
|u_k|=\frac{|v_k|}{a\sqrt{\GS}}\simeq \frac{1}{a\sqrt{2k c_{s+}\GS}}.
\end{align}
However, $ \mu_W^2 $ diverges at the zero point of $ \tilde\Theta $, implying that even in high $ k $ limit the WKB approximation does not hold around the zero point of $ \tilde\Theta $. For this reason, we divide the time history up to the bounce phase into two regions: one is the interval around the zero point of $ \tilde\Theta $ (denoted by $ t_0 $), where $ \tilde\Theta $ can be approximately described by a linear function of time, the other one is the remaining. We call these two regions respectively Region II and Region I, see Fig. \ref{fig:tildethetahregions}.

\begin{figure}
	\centering
	\includegraphics[width=\linewidth]{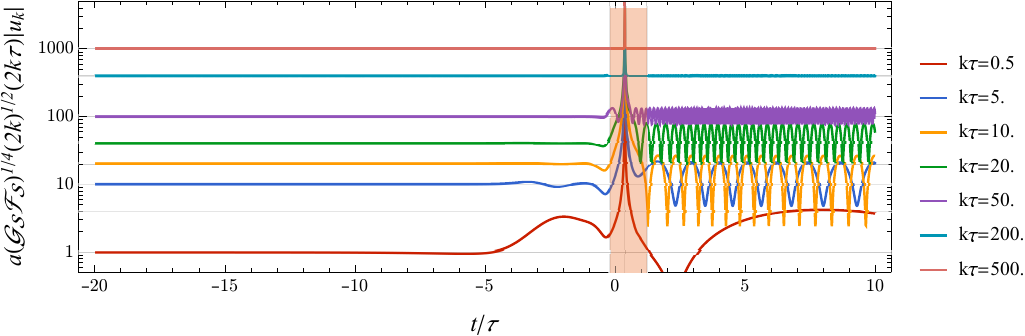}
	\caption{Evolution of $a(\GS\FS)^{1/4}(2k)^{1/2}(2k\tau)|u_k|$ for various wave number $ k $, confirming that \eqref{eq:RegionI-WKB-solution} holds in Region I. The region with orange background represents Region II. }
	\label{fig:evolution-uk-regionI}
\end{figure}

One can readily see from Fig. \ref{fig:tildethetahregions} that the WKB approximation holds in Region I for $ k\geq 5\tau^{-1} $ and thus the solution \eqref{eq:RegionI-WKB-solution} is valid, which can be numerically verified, see Fig.  \ref{fig:evolution-uk-regionI}. On the contrary, in Region II the WKB approximation does not hold and even the Mukhanov-Sasaki variable $ v_k $ becomes singular at $ t_0 $. In this region the original perturbation $ u_k $, however, remains finite and therefore we switch back to $ u_k $. Expanding the equation \eqref{eq:EOM-tildepsi} in Region II to the leading order, we obtain
\begin{align}
\ddot{u}_k-\frac{2}{t-t_0}\dot{u}_k+\frac{k^2}{a^2(t_0)}c_{s+}^2(t_0)u_k=0,
\end{align}
whose solution is given by
\begin{align}\label{eq:psitilde-around-t0}
u_k=C_0\left[-\omega_k(t-t_0)+i\right]e^{-i\left(\omega_k(t-t_0)+\varphi_0\right)},
\end{align}
with $ \omega_k\equiv \frac{c_{s+}(t_0)}{a(t_0)}k $. Here $ C_0 $ and $ \varphi_0 $ are integration constants and depend on the wave number $ k $.

\begin{figure}
	\centering
	\subfigure{\includegraphics[width=0.45\linewidth]{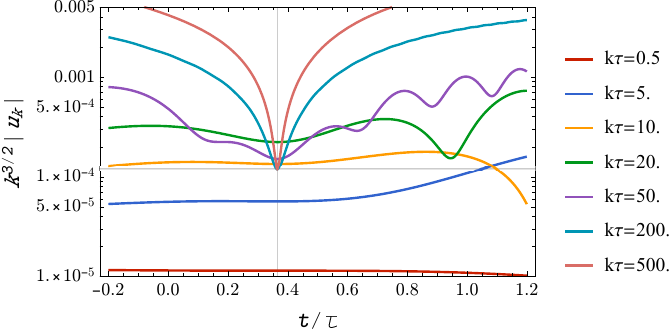}}
	\subfigure{\includegraphics[width=0.45\linewidth]{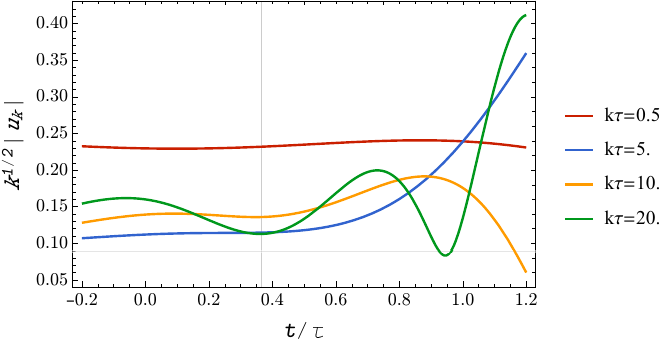}}
	\caption{Evolution of perturbations in Region II for various wave number $ k $. The left panel plots $ k^{\frac32}|u_k| $, confirming that \eqref{eq:psitilde-around-t0} holds in Region II and $ k^{\frac32}|u_k(t_0)| $ becomes a $ k $-independent constant $ C_1 $ for sufficiently high $ k $ (i.e. $ k\tau\geq 20 $). The right panel plots $ k^{\frac12}|u_k| $, showing that $ k^{\frac12}|u_k(t_0)| $ nearly does not depend on $ k $ for low $ k $ (i.e. $ 0.5\leq k\tau\leq 20 $). The gray vertical lines on the two panels indicate $ t_0 $. The gray horizontal line on the left panel represents the value of $ C_1 $, while on the right panel it indicates the value of $ C_2 $. }
	\label{fig:evolution-uk-regionII}
\end{figure}

The dependence of $ C_0 $ on $ k $ is not simple. Employing the continuity condition on the boundary between Region I and Region II, we find for $k$ satisfying $ \omega_k (t_0-t_-)\gg 1 $ (where $ t_- $ denotes the initial time of Region II )
\begin{align}\label{eq:C0-high-k}
C_0\simeq \frac{(2kc_{s+}(t_-)\GS(t_-))^{-\frac12}}{(t_0-t_-)a(t_-)\omega_k }\simeq C_1 k^{-\frac32}\;\text{ with }C_1\equiv \left(-\frac{\dot{\tilde\Theta}(t_0)}{2\Lambda(t_0)c_{s+}(t_0)}\right)^{\frac12},
\end{align}
where we have taken into account that the quantities like the sound speed squared and the scale factor do not change much in Region II. On the other hand, for $ k $ satisfying $ \omega_k(t_0-t_-)\lesssim \cO(1) $ we obtain
\begin{align}\label{eq:C0-low-k}
C_0\simeq\frac{ C_2}{a} k^{-\frac12}\;\text{ with }C_2\equiv \left.\frac{1}{\sqrt{2 c_{s+}\GS}}\right|_{t_-}.
\end{align}
The relations \eqref{eq:C0-high-k} and \eqref{eq:C0-low-k} can be numerically verified, see Fig. \ref{fig:evolution-uk-regionII}.

Inserting the solution \eqref{eq:psitilde-around-t0} into \eqref{eq:constraints-alpha-beta}, we obtain that the Fourier coefficients $ \alpha_k $ and $ \beta_k $ become 
\begin{align}\label{eq:regionII-alpha-beta}
\alpha_k\simeq
\frac{iC_0}{\dot{\tilde\Theta}(t_0)} \omega_k^2 e^{-i\left(\omega_k(t-t_0)+\varphi_0\right)},\quad \beta_k\simeq \frac{C_0\Lambda}{\dot{\tilde\Theta}(t_0)}\omega_k e^{-i\left(\omega_k(t-t_0)+\varphi_0\right)}.
\end{align}

\subsection{Comparison of third-order action with the second-order one}

In order to determine the strong-coupling scale, we compare the terms with the highest number of derivatives in the third-order and second-order actions \cite{Leblond:2008gg,Baumann:2011dt,Koehn:2015vvy}. After a long computation using Mathematica, we find that the constrained third-order action is given by
\begin{align}
\cS_{\phi}^{(3)}[\tilde\psi,\alpha,\beta]= \int d^3\mathbf x dt\;a^3\cL_\phi^{(3)},\label{eq:third-order-action-full}
\end{align}
where
\begin{align*}
\cL_\phi^{(3)}=\;& -\frac{1}{3}\tilde{\Omega } \alpha ^3 +\alpha ^2 \left(\frac{\beta _{,ii} }{12 a^2}\left(33 \Delta _0 \tilde{\Theta }-4 \tilde{\Xi }+42 \Delta _0 \dot{\Delta }_0 \mathcal{M}\right)+\tilde{\Xi }\dot{\tilde{\psi }} +3\hat{\Sigma } \tilde{\psi } \right.\\
&\left.-\frac{2 \tilde{\psi }_{,ii} }{a^2}\left(\Delta _0^2 F_2+2 \Delta _0 F_2+\Delta _1 F_2+4 \dot{\phi }^4 F_{2XX}+4 \Delta _0 \dot{\phi }^2 F_{2X}-2 \dot{\phi }^2 F_{2X}\right)\right)\\
&+\alpha  \left(\tilde{\psi } \left(\frac{\beta _{,ii} }{2 a^2}\left(2 \dot{\Delta }_0 \mathcal{M}-\tilde{\Theta }\right)-2\Lambda\frac{ \tilde{\psi }_{,ii} }{a^2}\right)+\frac{\Gamma }{2 a^4}\left(\beta _{,ii}{}^2-\beta _{,ij}{}^2\right)-\frac{\Lambda\tilde{\psi }_{,i}{}^2 }{a^2}+3 \Gamma  \dot{\tilde{\psi }}^2\right.\\
&\left.+\dot{\tilde{\psi }} \left(18(\tilde{\Theta }+\dot\Delta_0\cM) \tilde{\psi } +\frac{2 \beta _{,ii} }{a^2}\left(2 \dot{\phi }^2 \mathcal{M}_X+\Delta _0 \mathcal{M}-\mathcal{M}\right)\right)+\frac{\beta _{,i} \tilde{\psi }_{,i} }{2 a^2}\left(2 \dot{\Delta }_0 \mathcal{M}-\tilde{\Theta }\right)\right)\\
&-\frac{2\mathcal{M} \beta _{,ij} \beta _{,j} \tilde{\psi }_{,i} }{a^4}+\dot{\tilde{\psi }} \left(\frac{2\mathcal{M} \tilde{\psi } \beta _{,ii} }{a^2}+\frac{\alpha _{,i} \beta _{,i} }{a^2}\left(3 \dot{\phi }^2 \mathcal{T}+2 \Delta _0 \mathcal{M}\right)+\frac{2\mathcal{M} \beta _{,i} \tilde{\psi }_{,i} }{a^2}\right)\\
&+\tilde{\psi } \left(\frac{ \mathcal{M}}{2a^4}\left(\beta _{,ii}{}^2-\beta _{,ij}{}^2\right)+\frac{2 F_2 \tilde{\psi }_{,i}{}^2}{a^2}\right)-9\mathcal{M} \dot{\tilde{\psi }}^2 \tilde{\psi }.\numberthis\label{eq:third-order-Lag}
\end{align*}
The expressions for the coefficients $ \Gamma $, $ \tilde\Xi $, $ \tilde\Omega $ and $ \hat\Sigma $ are given in Appendix \ref{app:third-action-coeff}. As in deriving the second-order action, $ \tilde\psi^3 $ and $ \tilde\psi^2 \alpha $ terms were removed by using background equations of motion.

Note that in obtaining the above third-order action \eqref{eq:third-order-action-full} we did not use the constraint equation \eqref{eq:constraints-alpha-beta} and performed some integration by parts like $ C(t)\alpha^2\dot\alpha\to -\frac13\dot C(t)\alpha^3$. Strictly speaking, after performing the variable transformation $ \psi\to\tilde\psi $, the terms containing temporal derivative of $ \alpha $ and $ \beta $ should not appear in the action, otherwise the variational problem can not be well-defined. Therefore, suitable boundary terms should have been added to the original DHOST action \eqref{eq:DHOST-action} in order not to invoke the ill-posed variational problem. In the present work, we suppose that the original action has already this kind of boundary terms.

One might follow the standard procedure (see e.g. \cite{Maldacena:2002vr,Kobayashi:2011pc,Wang:2013eqj}) of substituting the non-dynamical variables $ \alpha $ and $ \beta $ by the constraint equations and performing further integrations by parts to remove some redundant operators in the third-order action \eqref{eq:third-order-action-full}. Indeed, in the inflationary context, the third-order action, which contains numerous terms as in \eqref{eq:third-order-action-full}, is greatly simplified by this procedure. Furthermore, all the terms become proportional to positive powers of the slow-roll parameter $ \epsilon_H\equiv -\dot H/H^2 $, which ensures the smallness of non-Gaussianity for the single-field slow-roll inflation theory \cite{Maldacena:2002vr}.

We point out that the above mentioned procedure is not suitable to estimating the strong-coupling scale in most bouncing scenarios, including the one considered in this work. The reason is twofold. First, this procedure leaves terms proportional to the equation of motion of the linear curvature perturbation and total derivative terms. Here, the total spatial derivatives can be safely dropped in cosmology, and total time derivatives without a time derivative of the fields do not contribute to correlation functions at any order in perturbation theory \cite{Braglia:2024xxx}. However, one cannot simply throw away total time derivatives including a time derivative of the field or the terms proportional to the linear perturbation equation, which should appear in pairs to ensure that the Lagrangian does not contain the second time-derivative of the field. While they can be ignored when computing the bispectrum of super-horizon modes at the tree-level \cite{Burrage:2011hd,Wang:2013eqj}, one of them makes non-trivial contribution to correlation functions for sub-horizon modes depending on which formalism (i.e. operator or path integral) is employed \cite{Braglia:2024xxx,Kawaguchi:2024xxx}.

Second, some integrations by parts generate terms with singular coefficients in most bouncing scenarios, including the one considered in this work. For instance, in the ghost condensate bouncing models studied in \cite{Koehn:2013ewd,Fertig:2016czu}, the non-dynamical variable $ \alpha $ is inversely proportional to $ H $ and thus temporal integrations by parts generate terms with coefficients proportional to $ -\dot H/H^2 $ in the third-order action, which become singular at the bounce point $ H=0 $. A similar issue arises for the model studied in this work since $ \alpha\propto \tilde\Theta^{-1} $ and $ \tilde\Theta $ crosses zero near the bounce point. In this regard, what they did in \cite{Koehn:2015vvy} to remove apparent singularities of the third-order action at the bounce point is to undo the integration by parts.

The terms with the highest number of derivatives in \eqref{eq:third-order-action-full} are
\begin{align}\label{eq:alphabetabeta}
\frac{\Gamma}{2a}\alpha(\beta_{,ii}^2-\beta_{,ij}^2),
\end{align}
which contain five derivatives. Due to the structure of $ \alpha $ and $ \beta $, it is convenient to deal with \eqref{eq:alphabetabeta} for two separate cases.
 
\textbf{Away from the zero point of $ \tilde\Theta $ (Region I in Fig. \ref{fig:tildethetahregions})} In this case the relevant part of the terms \eqref{eq:alphabetabeta} is (see e.g. Appendix C of \cite{DeFelice:2011zh})
\begin{align}\label{eq:psiddpsiddpsi}
\frac{\Gamma\Lambda^2}{2a\tilde\Theta^3}\dot{\tilde\psi}\left(\tilde\psi_{,ii}^2-\tilde\psi_{,ij}^2\right)\dot=&-\frac13\frac{d}{dt}\left(\frac{\Gamma\Lambda^2}{2a\tilde\Theta^3}\right)\tilde\psi\left(\tilde\psi_{,ii}^2-\tilde\psi_{,ij}^2\right),
\end{align}
where the notation $ \dot= $ refers to equality up to a total derivative. The RHS of \eqref{eq:psiddpsiddpsi} contains four derivatives, which together with other terms containing four derivatives in \eqref{eq:third-order-action-full} 
\begin{align}
&a^3\left[\frac{\mathcal{M}}{2a^4}\tilde\psi\left(\beta _{,ii}{}^2-\beta_{,ij}{}^2\right)-\frac{2\mathcal{M} \beta _{,ij} \beta _{,j} \tilde{\psi }_{,i} }{a^4}+\alpha\dot{\tilde\psi}\frac{2 \beta _{,ii} }{a^2}\left(2 \dot{\phi }^2 \mathcal{M}_X+\Delta _0 \mathcal{M}-\mathcal{M}\right) \right]\simeq\nonumber\\
&\simeq \frac{\mathcal{M}\Lambda^2}{2a\tilde\Theta^2}\tilde\psi\left(\tilde\psi_{,ii}{}^2-\tilde\psi_{,ij}{}^2\right)-\frac{2\mathcal{M}\Lambda^2 \tilde\psi_{,ij} \tilde\psi_{,j} \tilde{\psi }_{,i} }{a\tilde\Theta^2}\nonumber\\
&\hskip10em-\dot{\tilde\psi}^2\frac{2a\Lambda \tilde\psi_{,ii} }{\tilde\Theta^2}\left(2 \dot{\phi }^2 \mathcal{M}_X+\Delta _0 \mathcal{M}-\mathcal{M}\right),
\end{align}
form the relevant terms in high $ k $ limit in the third-order action. The ratio of these to the second-order action then becomes
\begin{align}
\frac{\cL^{(3)}_\phi}{\cL^{(2)}_\phi}\sim &\left[\frac13\left|\frac{d}{dt}\left(\frac{\Gamma\Lambda^2}{a\tilde\Theta^3}\right)\right|+3\left|\frac{\cM\Lambda^2}{a\tilde\Theta^2}\right|+\left|\frac{2\Lambda c_{s+}^2}{a\tilde\Theta^2 }\left(2 \dot{\phi }^2 \mathcal{M}_X+\Delta _0 \mathcal{M}-\mathcal{M}\right)\right| \right]\frac{k^3 c_{s+}^{\frac12}}{\sqrt 2 a^2\FS^{\frac32}},\label{eq:ratio-L3-L2}
\end{align}
where we have taken into account $ |\tilde{\psi}|\sim|k^{3/2}\tilde{\psi}_k| \sim \frac{k}{a\sqrt{2c_{s+}\GS}} $, see \eqref{eq:RegionI-WKB-solution} and \cite{Dehghani:2025xxx}. Therefore, the ratio $ \cL^{(3)}_\phi/\cL^{(2)}_\phi $ scales as $ k^3 $ in Region I. When this ratio becomes $ \cO(1) $, the model gets strongly coupled, which occurs at the scale 
\begin{align}\label{eq:Lambdas-RegionI}
\frac{\Lambda_s}{a}\propto \tilde\Theta^{\frac23}\propto E_{\text{back}}^{\frac23}.
\end{align}
Since $ E_{\text{back}}\lesssim \cO(1)\tau^{-1}\simeq 10^{-4} $, we find the strong-coupling scale $ \frac{\Lambda_s}{a}\gtrsim 10E_{\text{back}}\gg E_{\text{back}} $. 

During the ekpyrotic contracting phase, one can estimate the strong coupling scale more precisely, namely
\begin{align}\label{eq:strong-scale-ekpyrotic}
\frac{\Lambda_s}{a}\sim \left(\frac{3\epsilon^{\frac32}}{3\epsilon+16} H^2\right)^{\frac13},
\end{align}
which is much higher than the background energy scale $ E_{\text{back}}\simeq \epsilon^{\frac12}|H| $, but is much lower than the already known result $ \sqrt{\frac{2}{\epsilon}} $ \cite{Koehn:2015vvy}.

This discrepancy stems from the fact that in \cite{Koehn:2015vvy} the strong-coupling scale was evaluated in the spatially-flat gauge where the curvature perturbation is fixed to be zero. The curvature perturbation $ \psi $ in the unitary gauge is related to the field perturbation $ \delta\varphi $ in the spatially-flat gauge by the non-linear gauge transformation\footnote{Here we denote the canonically normalized scalar field by $ \varphi $. Note that during the ekpyrotic contracting phase, we have $ -H/\dot{\bar\varphi}\simeq 1/\sqrt{2\epsilon} $.} \cite{Maldacena:2002vr,Koyama:2010xj}
\begin{align}
\psi=\psi_n+\cZ(\psi_n),
\end{align}
where $ \psi_n=-H\delta\varphi/\dot{\bar\varphi} $ and 
\begin{align}
\cZ(\psi)=\;&\frac{\eta_H}{4}\psi^2+\frac{\psi\dot\psi}{H}+\frac{1}{4a^2H^2}\left[-\psi_{,i}^2+\partial^{-2}(\partial_i\partial_j(\psi_{,i}\psi_{,j})) \right]\nonumber\\
&\hskip3em+\frac{\epsilon_H}{2H}\left[\psi_{,i}\partial_i(\partial^{-2}\dot\psi)-\partial^{-2}(\partial_i\partial_j(\psi_{,i}\partial_j(\partial^{-2}\dot\psi)))\right],
\end{align}
with $ \epsilon_H\equiv -\dot H/H^2 $ and $ \eta_H=\frac{\dot\epsilon_H}{H\epsilon_H} $. Strongly-coupled terms would therefore be generated by the gauge transformation from the spatially flat gauge to the unitary one if $ \cZ(\psi_n)/\psi_n\sim \cO(1) $. One can readily see that this occurs at the scale $ \left(\sqrt{2\epsilon}H^2\right)^{\frac13} $, which is close to \eqref{eq:strong-scale-ekpyrotic}, taking $ \epsilon\gg 1 $ into account.

\textbf{Around the zero point of $ \tilde\Theta $ (Region II in Fig. \ref{fig:tildethetahregions})} In this case the terms \eqref{eq:alphabetabeta} can not be written as \eqref{eq:psiddpsiddpsi}, due to the singular behavior of $ \tilde\Theta $. It follows from \eqref{eq:regionII-alpha-beta} along with \eqref{eq:C0-high-k} and \eqref{eq:C0-low-k} that the ratio of the third-order action to the second-order one in this region becomes
\begin{align}
\cL^{(3)}_\phi/\cL^{(2)}_\phi\sim \frac{\Gamma}{2a^4\GS\tilde\Theta^2}\frac{k^4\alpha\beta^2}{\alpha^2}\sim\begin{cases}
\frac{C_1\Gamma}{2a^4\GS\tilde\Theta^2}k^4 & \text{for } k\tau\geq 20,\\
\frac{C_2\Gamma}{2a^3\GS\tilde\Theta^2}k^3 & \text{for } k\tau\leq 20.
\end{cases}
\end{align}
Hence, the ratio $ \cL^{(3)}_\phi/\cL^{(2)}_\phi $ increases more rapidly in high $ k $ regime, which implies that we have only to consider the case of $ k\tau\leq 20 $. Estimating the strong-coupling scale in this region, we obtain
\begin{align}\label{eq:Lambdas-RegionII}
\frac{\Lambda_s}{a}\sim \left(\frac{2\GS\tilde\Theta^2}{C_2\Gamma}\right)^{\frac13}\sim\tilde\Theta(t_0)^{\frac23} \sim E_{\text{back}}^{\frac23}\sim 20E_{\text{back}}\gg E_{\text{back}}.
\end{align}

\subsection{Strong-coupling scale through three-point function}

The more robust estimation of the strong-coupling scale is obtained through the 3-point function \cite{Dehghani:2025xxx}
\begin{align}
&\braket{\tilde\psi_{\mathbf k_1}(t_e)\tilde\psi_{\mathbf k_2}(t_e)\tilde\psi_{\mathbf k_3}(t_e)}\equiv (2\pi)^3\delta^{(3)}(\mathbf k_1+\mathbf k_2+\mathbf k_3)B(\mathbf k_1,\mathbf k_2,\mathbf k_3).
\end{align}
In this work we consider only one-vertex order, at which the bispectrum can be computed by (see e.g. \cite{Wang:2013eqj} for a review)
\begin{align}
&\braket{\tilde\psi_{\mathbf k_1}(t_e)\tilde\psi_{\mathbf k_2}(t_e)\tilde\psi_{\mathbf k_3}(t_e)}=\nonumber\\
&\hskip5em-2\text{Im}\braket{\tilde\psi_{\mathbf k_1}(t_e)\tilde\psi_{\mathbf k_2}(t_e)\tilde\psi_{\mathbf k_3}(t_e)\int_{-\infty(1+i\epsilon)}^{t_e}dt\int d^3\mathbf x\;a^3\cL^{(3)}_\phi}.
\end{align}

The non-Gaussianity parameter $ f_{NL} $ is defined by 
\begin{align}
f_{NL}(\mathbf k_1,\mathbf k_2,\mathbf k_3)=\frac{B(\mathbf k_1,\mathbf k_2,\mathbf k_3)}{2\left(P(\mathbf k_1)P(\mathbf k_2)+P(\mathbf k_2)P(\mathbf k_3)+P(\mathbf k_3)P(\mathbf k_1)\right)},
\end{align}
where
\begin{align}
\braket{\tilde\psi_{\mathbf k}\tilde\psi_{\mathbf k'}}=(2\pi)^3 \delta^{(3)}(\mathbf k+\mathbf k')P(k),\quad P(k)=|u_k|^2=\frac{2\pi^2}{k^3}\cP(k)
\end{align}
The model then becomes strongly coupled when the quantity
\begin{align}\label{eq:fNL-psi}
\left|f_{NL}(k,k,k)\tilde\psi\right|\approx \frac{k^6 B(k,k,k)}{24\pi^4\cP^{3/2}(k)}
\end{align}
becomes $ \cO(1) $ \cite{Dehghani:2025xxx}. 

\begin{figure}
	\centering
	\includegraphics[width=\linewidth]{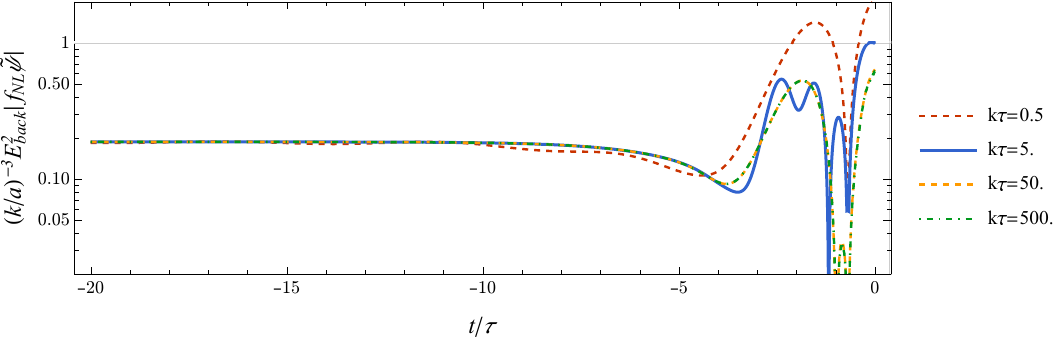}
	\caption{Plot of $(\frac{k}{a})^{-3}E_{\text{back}}^2|f_{NL}\tilde\psi|$ for several $k$. The plot shows that $(\frac{k}{a})^{-3}E_{\text{back}}^2|f_{NL}\tilde\psi|\lesssim \cO(1)$ for a wide range of $ k $. }
	\label{fig:fnlscaledregioni}
\end{figure}

We have numerically evaluated \eqref{eq:fNL-psi} for several $ k $. The result for Region I is shown in Fig. \ref{fig:fnlscaledregioni}. One can readily see that in Region I
\begin{align}
|f_{NL}\tilde\psi|\simeq \cO(1)\times \left(\frac{k}{a}\right)^3E_{\text{back}}^2,
\end{align}
and thus the strong-coupling scale is
\begin{align}
\frac{\Lambda_s}{a}\sim E_{\text{back}}^{\frac23},
\end{align}
which is consistent with \eqref{eq:Lambdas-RegionI}.

\begin{figure}
	\centering
	\includegraphics[width=\linewidth]{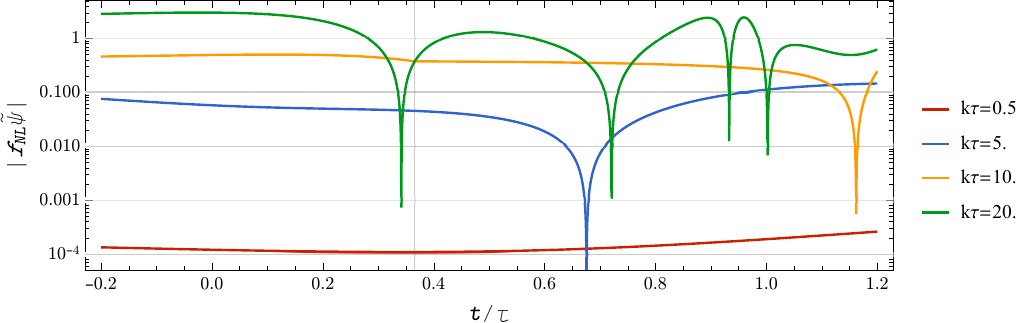}
	\caption{Plot of $|f_{NL}\tilde\psi|$ in Region II for several $ k $. The plot shows that $|f_{NL}\tilde\psi|$ scales roughly as $ k^3 $ for $ k\tau\leq 20 $ and the model becomes strongly coupled at $ k\tau\simeq 10 $.}
	\label{fig:fnllowkregionii}
\end{figure}

On the other hand, the dependence of $ |f_{NL}\tilde\psi| $ on $ k $ in Region II is not simple as in Region I, see Fig. \ref{fig:fnllowkregionii}, from which one can roughly say that $ |f_{NL}\tilde\psi|\propto k^3 $ for $ k\tau\leq 20 $. Fig. \ref{fig:fnllowkregionii} also shows that the model becomes strongly coupled at the scale $ k\tau\simeq 10 $, implying that the strong-coupling scale is
\begin{align}
\frac{\Lambda_s}{a}\simeq E_{\text{back}}^{\frac23}\simeq 20E_{\text{back}},
\end{align}
which is again consistent with the previous estimation \eqref{eq:Lambdas-RegionII}.

\begin{figure}
	\centering
	\includegraphics[width=0.7\linewidth]{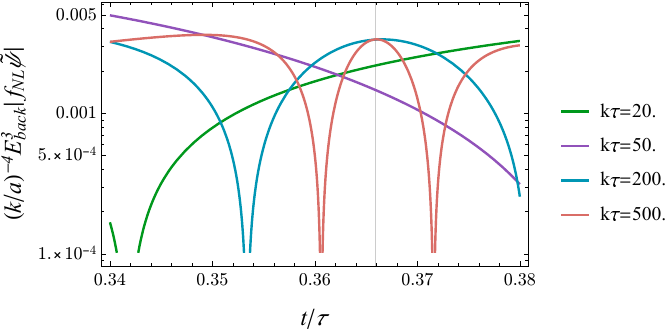}
	\caption{Plot of $(k/a)^{-4}E_{\text{back}}^{-3}|f_{NL}\tilde\psi|$ for $ k\tau\geq 20 $, showing that $|f_{NL}\tilde\psi|$ scales as $ k^4 $ around $ t_0 $ for $ k\geq 20\tau^{-1} $.}
	\label{fig:fnlhighkregionii}
\end{figure}

In the previous subsection we have seen that the ratio $ \cL_\phi^{(3)}/\cL_\phi^{(2)} $ scales as $ k^4 $ for high $ k $, which should be reflected in the scaling behavior of $ |f_{NL}\tilde\psi| $ around $ t_0 $. To see this, we plot $ \left(\frac{k}{a}\right)^{-4}E_{\text{back}}^3|f_{NL}\tilde\psi| $ around $ t_0 $ in Fig. \ref{fig:fnlhighkregionii}, which confirms that $ |f_{NL}\tilde\psi| $ scales as $ k^4 $ for $ k\tau\geq 20 $.  We emphasize that this analysis is in fact irrelevant to estimation of the strong-coupling scale but is important to verify the role of the term $ \frac{\Gamma}{2a}\alpha(\beta_{,ii}^2-\beta_{,ij}^2) $.

We have not dealt with kinetically driven expanding phase after the bounce, but one can easily see that the strong-coupling scale in that phase is essentially the same as in the ekpyrotic contracting phase. In summary, we find that the strong-coupling scale $ \frac{\Lambda_s}{a} $ at all times is order of $ E_{\text{back}}^{\frac23} $, which is much higher than the characteristic background energy scale $ E_{\text{back}} $. We thus conclude that the bouncing background of our model is safe from strong-coupling issue and the model is self-consistent.

\section{Conclusion}\label{sec:conclusion}
In this work, we have constructed a two-field DHOST bouncing model which indeed has no pathologies and is compatible with observations, focusing on non-Gaussianity and strong-coupling problems arising at the non-linear level. We first refined  the two-field DHOST bouncing model proposed in \cite{An:2025xxx} without altering its viability at the linear level, such that all the higher-order terms of the action are suppressed outside the bounce phase, which makes the local non-Gaussianity of the model well under control. It was then demonstrated that there exists a region of parameter space where the local non-Gaussianity agrees with the observations.

We also showed that the refined model can successfully circumvent the strong coupling problem. To be concrete, it was found that the strong-coupling scale of the model is order of $ \left(M_{pl}E_{\text{back}}^{2}\right)^{\frac13} $ which is well above the characteristic background energy scale $ E_{\text{back}}\lesssim \tau^{-1}M_{pl} $ (where $ \tau\sim\cO(10^{4}-10^5) $) at all times.\footnote{Here we restored the reduced Planck mass $M_{pl}$ for easy comparison of the strong coupling scale to the Planck scale.} This implies that the model is weakly-coupled and the classical description of the background bouncing solution is robust and trustworthy.
 An important finding here is that the strong-coupling scale during the ekpyrotic contracting phase in the unitary gauge is much lower than in the spatially-flat gauge that is known to be close to the Planck scale. This is in fact a natural consequence of the non-linear gauge transformation from the unitary gauge to the spatially-flat gauge.

In summary, the model presented in this work, refined version of \cite{An:2025xxx}, is fully viable in the sense that not only it is weakly-coupled, stable and non-superluminal but also it can predict nearly scale-invariant curvature perturbations with small non-Gaussianity compatible with observations.

\appendix

\section{Perturbation expansion of the action}\label{app:perturbation-expansion}

In this appendix we perform the perturbation expansion for action of the exceptional subclass of DHOST Ia coupled to a luminal scalar field up to the second order. We focus on the scalar perturbations, while the second-order action for tensor perturbations can be found elsewhere \cite{Mironov:2018oec,Mironov:2024xxx}.

We begin with the ADM decomposition of the action $ \cS_{\text{full}} $, which is given by \cite{Langlois:2017mxy}
\begin{align}\label{eq:action-ADM-full}
\cS_{\text{full}}=\int d^3\mathbf x\; dt\; N\sqrt{\gamma}\;(\cL_\phi+\cL_\chi),
\end{align}
with
\begin{align*}
\cL_\phi=&\;F+F_2R[\gamma]-2F_{2\phi}vK+(F_2+A_1 v^2)K_{ij}K^{ij}-(F_2-A_2 v^2)K^2\\
&+\left[A_1+A_2-(A_3+A_4)v^2+A_5 v^4\right]V^2+(4F_{2X}+2A_2-A_3 v^2)vKV\\
&+\left[4F_{2X}v^2-(2A_1-A_4 v^2)v^2\right]\frac{\partial_i N\partial^i N}{N^2},\numberthis\label{eq:Lag-phi}\\
\cL_\chi=&\;\frac{Q}{N^2}(\dot\chi-N^i\partial_i\chi)^2 -Q\gamma^{ij}\partial_i\chi\partial_j\chi-W(\phi,\chi),\numberthis\label{eq:Lag-chi}
\end{align*}
where
\begin{subequations}
	\begin{align}
	& K_{ij}=\frac{1}{2N}(\dot \gamma_{ij}-\nabla_i N_j-\nabla_j N_i),\quad K\equiv \gamma^{ij}K_{ij},\\
	& v\equiv \frac{1}{N}\left(\dot\phi-N^i\partial_i\phi\right)=\frac{\dot\phi}{N}, \quad V\equiv \frac{1}{N}\left(\dot v-N^i\partial_i v\right)=\frac{1}{N}\left(\dot v+v\frac{N^i\partial_i N}{N}\right),
	\end{align}
\end{subequations}
and the notation $ \nabla $ indicates the covariant derivative with respect to the induced metric $ \gamma_{ij} $. Introducing the following auxiliary functions
\begin{subequations}\label{eq:definition-cM-cB}
	\begin{align}
	& \cM\equiv 2(F_2-A_1 X),\\
	& \cB\equiv (A_3+A_4)X+A_5 X^2=\frac{3X\cT^2}{4\cM},\\
	& \cT\equiv 4F_{2X}-2A_1+A_3 X,\\
	& \cJ\equiv -4F_{2X}X+2A_1 X+A_4 X^2=2(2X F_{2X}-F_2)\frac{X\cT}{\cM}-2F_2\left(\frac{X\cT}{2\cM}\right)^2,
	\end{align}
\end{subequations}
the Lagrangian $ \cL_\phi $ can be written as
\begin{align}
\cL_\phi=F+F_2 R[\gamma]-2F_{2\phi}vK+\frac12\cM(K_{ij}K^{ij}-K^2)+\cB V^2+\cT vKV+\cJ\frac{\partial_i N\partial^i N}{N^2}.\label{eq:L-DHOST}
\end{align}

Now we insert perturbed field \eqref{eq:metric-perturbation} and \eqref{eq:chi-perturbation} into the ADM decomposed Lagrangian \eqref{eq:L-DHOST} to perform perturbation expansion. At linear order, it gives background equations of motion, namely
\begin{subequations}
	\begin{align*}
	& 2 \dot{\phi } \dddot{\phi } \mathcal{B}+\ddot{\phi }^2 \left(-2 \dot{\phi }^2 \mathcal{B}_X-\mathcal{B}\right)+\ddot{\phi } \left(3 H \dot{\phi } (2 \mathcal{B}-\mathcal{T})+2 \dot{\phi }^2 \mathcal{B}_{\phi }\right)\\
	&\hskip2em+3 H \dot{\phi }^3 (\cT_{\phi }-4 F_{2\phi} X)+\dot{\phi }^2 \left(-6 H^2 \cM_X+9 H^2 \mathcal{T}+3 \dot{H} \mathcal{T}+2 F_X\right)\\
	&\hskip2em+6 H \dot{\phi } F_{2\phi }+3\cM H^2 +F-Q\dot\chi^2-W=0\numberthis,\\
	&-\dot{\phi } \dddot{\phi } \mathcal{T}+ \ddot{\phi }^2 \left(2 \dot{\phi }^2 \mathcal{T}_X-\mathcal{T}+\mathcal{B}\right)+\ddot{\phi } \left(-4 H \dot{\phi } \mathcal{M}_X+\dot{\phi }^2 (-4 F_{2\phi X}-\mathcal{T}_{\phi })+2 F_{2\phi }\right)\\
	&\hskip2em+2 \dot{\phi }^2 F_{2\phi \phi }+2 H \dot{\phi } \mathcal{M}_{\phi }+2 \dot{H} \mathcal{M}+3 H^2 \mathcal{M}+F+Q\dot\chi^2-W=0,\numberthis\\
	& Q\ddot\chi+\left(3HQ+\dot Q\right)\dot\chi+\frac12\frac{\partial W}{\partial\chi}=0.\numberthis
	\end{align*}\label{eq:bg-eqns}
\end{subequations}

Before moving to the higher order, we first observe that due to the DHOST Ia conditions \eqref{eq:DHOST-I}, \eqref{eq:DHOST-Ia1} and \eqref{eq:DHOST-Ia2} the terms quadratic in $ \dot\psi $ and $ \dot\alpha $ in the full perturbed Lagrangian $ \cL_\phi $ form a perfect square, namely
\begin{align*}
\cL_{\phi}=&-\frac{3\cM}{N^2}\left(\dot\psi-\Delta\dot\alpha\right)^2+\cdots
\end{align*}
where
\begin{align*}
\Delta\equiv \frac{X\cT}{2N\cM},
\end{align*}
implying that one of $ \psi $ and $ \alpha $ is non-dynamical. In order to remove the kinetic mixing between $ \psi $ and $ \alpha $, we introduce a new variable $ \tilde\psi $ such that $ \dot\alpha $ in the perfect square is absorbed away. The new variable $ \tilde\psi $ should be defined by
\begin{align}
\dtpsi=\dot\psi-\Delta\dot\alpha,
\end{align}
which, however, can not be integrated exactly. Since our purpose is to perform the perturbation expansion of the action up to the third order, we integrate  the above relation with respect to $ t $ up to the second order in $ \alpha $ to obtain
\begin{align}
\tilde\psi=\psi-\Delta_0\alpha-\frac12\Delta_1\alpha^2+\cO(\alpha^3),
\end{align}
where
\begin{align}
\Delta=\Delta_0+\Delta_1 \alpha+\cO(\alpha^2)
\end{align}
with
\begin{align}\label{eq:Delta0-1}
\Delta_0=-\frac{\cT\dot\phi^2}{2\cM},\quad  \Delta_1=\frac{\dot\phi^2}{2\cM^2}\left(3\cM\cT+2\dot\phi^2\cT\cM_X-2\dot\phi^2\cM\cT_X \right).
\end{align}
In the above equation and in the following all the quantities are evaluated at the background level, e.g. $ X=-\dot\phi^2 $.

The second-order action for $ \phi $ sector is given by
\begin{align}
\cS_{\phi}^{(2)}[\tilde\psi,\alpha,\beta]=&\;\int d^3\mathbf x dt\; a^3\Bigg[-3\cM\dtpsi^2+2 F_2\frac{\tilde\psi_{,i}\tilde\psi_{,i}}{a^2}+6\tilde\Theta\dtpsi\alpha\nonumber\\
&\hskip7em +2(\cM\dtpsi-\tilde\Theta\alpha)\frac{\beta_{,ii}}{a^2}+\tilde\Sigma\alpha^2-2\Lambda\alpha\frac{\tilde\psi_{,ii}}{a^2}\Bigg],\label{eq:second-order-phi-sector-constrained}
\end{align}
where
\begin{subequations}\label{eq:functions-general}
	\begin{align}
	& \Lambda=2(1+\Delta_0) F_2+4\dot\phi^2  F_{2X},\\
	& \tilde\Theta=\Theta-\cM\dot\Delta_0,\\
	& \tilde\Sigma=\Sigma+3 \Theta  \dot\Delta_0-3 \Delta_0 \dot\Theta-\frac32 \cM \dot\Delta_0^2-9H \Delta_0 \Theta,\\
	& \Theta=\ddot\phi \left(\dot\phi^3 \cT_X+\dot\phi (\cB-\frac 32 \cT)\right)-2 \dot\phi^3 F_{2\phi X}\nonumber\\
	&\hskip5em+\frac{1}{2} \dot\phi^2 \left(3\cT H -4\cM_X H \right)+\dot\phi F_{2\phi }+\cM H ,\\
	&\Sigma=\dddot\phi \dot\phi \left(2 \dot\phi^2 \cB_X-4 \cB \right)+\ddot\phi^2 \left(5 \dot\phi^2 \cB_X-2 \dot\phi^4 \cB_{XX}+2 \cB\right)\nonumber\\
	&\hskip2em+\ddot\phi \left(3H \dot\phi^3 \left(2 \cB_X- \cT_X\right)-6H \dot\phi (2 \cB- \cT)-4 \dot\phi^2 \cB_{\phi }+2 \dot\phi^4 \cB_{\phi X}\right)\nonumber\\
	&\hskip2em+3H \dot\phi^5 \left( \cT_{\phi X}-4 F_{2\phi XX}\right)+\dot\phi^4 \left(3 \dot H \cT_X+9 H^2 \cT_X-12 H^2 \cM_{XX}+2 P_{XX}\right)\nonumber\\
	&\hskip2em-6H \dot\phi^3 \left( \cT_{\phi }-5 F_{2\phi X}\right)+\dot\phi^2 \left(15\cM_X H^2 -6 \cT \dot H-18 H^2 \cT-P_X\right)\nonumber\\
	&\hskip2em-6 H \dot\phi F_{2\phi}-3\cM H^2.
	\end{align}
\end{subequations}
On the other hand, the second-order action for $ \chi $ sector is simply given by
\begin{align}
\cS^{(2)}_\chi[\delta\chi,\alpha,\beta]=&\int d^3\mathbf x dt \;a^3\left[Q\left(\dot{\bar\chi}^2\alpha^2-2\dot{\bar\chi} \alpha\dot{\delta\chi}+2\dot{\bar\chi}\delta\chi\frac{\beta_{,ii}}{a^2}+\dot{\delta\chi}^2-\frac{\delta\chi_{,i}\delta\chi_{,i}}{a^2}-6\dot{\bar\chi}\dot\psi\delta\chi \right)\right.\nonumber\\
&\hskip6em \left.-W_{\chi}\alpha\delta\chi-\frac12 W_{\chi\chi}\delta\chi^2\right]\label{eq:second-order-chi-sector-constrained}
\end{align}
In obtaining \eqref{eq:second-order-phi-sector-constrained} and \eqref{eq:second-order-chi-sector-constrained} we implemented some integration by parts and took into account the background equations of motion \eqref{eq:bg-eqns} to remove the terms proportional to $ \psi^2 $ and $ \psi\alpha $.

Taking functional derivatives for the full second-order action $ \cS_\phi^{(2)}+\cS_\chi^{(2)} $ with respect to $ \beta $ and $ \alpha $, we obtain algebraic constraints for $ \alpha $ and $ \beta $, namely
\begin{subequations}\label{eq:constraints-full}
	\begin{align}
	& \alpha=\frac{\cM\dtpsi+Q\dot{\bar\chi}\delta\chi}{\tilde\Theta},\\
	& \frac{1}{a^2}\beta_{,ii}=3\dtpsi+\frac{1}{\tilde\Theta}\left(\tilde\Sigma\alpha-\Lambda\frac{\tilde\psi_{,ii}}{a^2}+Q\dot{\bar\chi}^2\alpha-\dot{\bar\chi}\dot{\delta\chi}-\frac12 W_\chi\delta\chi\right).
	\end{align}
\end{subequations}
Substituting these constraints into the second-order action $ \cS_\phi^{(2)}+\cS_\chi^{(2)} $, we obtain the scalar part of unconstrained second-order action \eqref{eq:second-order-full-action}.

\section{Large-scale evolution of the perturbations up to the second order}\label{app:LS-evolution}

Large-scale equations for the adiabatic and entropic perturbations are well-formulated in the covariant formalism (see e.g. \cite{Langlois:2006vv} and references therein). For the two scalar fields $\phi^I $ ($ I,J=1,2 $) with the field space metric $ G_{IJ} $, we define the adiabatic perturbation $ \delta\sigma $ and the entropic perturbation $ \delta s $ up to the second order by \cite{RenauxPetel:2008gi,Fertig:2015ola}
\begin{subequations}
\begin{align}
& \delta\sigma\equiv \bar e_{\sigma I}\delta\phi^I,\quad \delta s\equiv \bar e_{s I}\delta\phi^I,\\
& \delta\sigma^{(2)}\equiv \bar e_{\sigma I}\delta\phi^{I(2)}+\frac12\bar e_{\sigma J}\bar\Gamma^J_{IK}(\bar e^K_\sigma \bar e^I_\sigma \delta\sigma^2+\bar e^K_s\bar e^I_s\delta s^2+2\bar e^K_\sigma \bar e^I_s\delta\sigma\delta s)+\frac{1}{2\dot{\bar\sigma}}\delta s\dot{\delta s},\\
& \delta s^{(2)}\equiv \bar e_{s I}\delta\phi^{I(2)}+\frac12\bar e_{s J}\bar\Gamma^J_{IK}(\bar e^K_\sigma \bar e^I_\sigma \delta\sigma^2+\bar e^K_s\bar e^I_s\delta s^2+2\bar e^K_\sigma \bar e^I_s\delta\sigma\delta s)\nonumber\\
&\hskip25em-\frac{\delta\sigma}{\dot{\bar\sigma}}\left(\dot{\delta s}+\frac{\dot{\bar\theta}}{2}\delta\sigma\right),
\end{align}
\end{subequations}
where the superscript $ (2) $ stands for the second-order perturbation, $ \Gamma^J_{IK} $ is the Christoffel symbol associated with the field space metric $ G_{IJ} $, and
\begin{subequations}
\begin{align}
	& e^I_\sigma\equiv \frac{\cL_u\phi^I}{\dot\sigma},\quad G_{IJ}e^I_s e^J_s=1,\quad G_{IJ}e^I_s e^J_\sigma=0,\quad \dot\sigma\equiv \sqrt{G_{IJ}(\cL_u\phi^I)( \cL_u\phi^J)},\\
& \cL_u e^I_\sigma\equiv \dot\theta e^I_s,\quad \cL_u e^I_s\equiv -\dot\theta e^I_\sigma,
\end{align}
\end{subequations}
with the overbar notation representing the background value and the notation $ \cL_u $ denoting the Lie derivative with respect to a time-like vector $ u^\mu=\frac{1}{N}(1,0,0,0) $.

Our aim is to provide the large-scale evolution of the gauge-invariant curvature perturbation $ \zeta $, defined by \cite{Langlois:2006vv,Langlois:2010vx}
\begin{subequations}
\begin{align}
& \zeta\equiv -\psi-\frac{H}{\dot{\bar\rho}}\delta\rho,\\
& \zeta^{(2)}\equiv -\psi^{(2)}-\psi^2-\frac{H}{\dot{\bar\rho}}\delta\rho^{(2)}-\frac{\delta\rho}{\dot{\bar\rho}}\left[\dot\zeta+\frac12\frac{d}{dt}\left(\frac{H}{\dot{\bar\rho}}\right)\delta\rho\right],
\end{align}
\end{subequations}
where $ \rho $ stands for the energy density. In the comoving gauge the evolution equation of the entropic fluctuation $ \delta s $ and the curvature perturbation $ \zeta $ can be expressed on large-scales as \cite{Fertig:2015ola}
\begin{subequations}\label{eq:perturbation-EOM}
	\begin{align}
	& \ddot{\delta s}+3H \dot{\delta s}+\left(V_{;ss}+3\dot\theta^2+\dot\sigma^2 e^I_s e^J_s e^K_\sigma e^L_\sigma R_{IKJL}\right)\delta s= 0,\\
	& \ddot{\delta s^{(2)}}+3H\dot{\delta s^{(2)}}+\left(V_{;ss}+3\theta'^2+\dot\sigma^2 e^I_s e^J_s e^K_\sigma e^L_\sigma R_{IKJL}\right)\delta s^{(2)}=\nonumber\\
	&\hskip2em -\frac{\dot\theta}{\dot\sigma}\dot{\delta s}^2-\frac{2}{\dot\sigma}\left[\ddot\theta+\frac{V_{,\sigma}\dot\theta}{\dot\sigma}-\frac32 H\dot\theta \right]\delta s\dot{\delta s}+\left[-\frac12 V_{;sss}+\frac{5V_{;ss}\dot\theta}{\dot\sigma}+\frac{9\dot\theta^3}{\dot\sigma} \right.\nonumber\\
	&\hskip2em\left.+e^I_s e^J_s e^K_\sigma e^L_\sigma\left(\dot\sigma\dot\theta R_{IKJL}-\frac12 \dot\sigma^2 e^N_s \cD_N R_{IKJL} \right)\right]\delta s^2,\\
	& \dot\zeta= -\frac{2H\dot\theta}{\dot\sigma}\delta s,\\
	& \dot{\zeta^{(2)}}= \frac{2H}{\dot\sigma^2}\left[-\dot\sigma \dot\theta\delta s^{(2)}-\frac{V_{,\sigma}}{2\dot\sigma}\delta s\dot{\delta s}+\left(\frac12 V_{;ss}+2\dot\theta^2\right)\delta s^2 \right],
	\end{align}
\end{subequations}
where $ R_{IJKL} $ and $ \cD_N $ are respectively the Riemann tensor and covariant derivative associated with the metric $ G_{IJ} $. And we introduced the notation
\begin{subequations}
	\begin{align}
	& V_{,\sigma}=e^I_\sigma V_{,I},\quad V_{,s}=e^I_s V_{,I},\quad V_{;ss}=e^I_s e^J_s \cD_I\cD_J V,\quad V_{;sss}=e^I_s e^J_s e^K_s \cD_I \cD_J \cD_K V.
	\end{align}
\end{subequations}

\section{Third-order action coefficients}\label{app:third-action-coeff}

In this appendix we provide the general expressions for the coefficients in the third-order action \eqref{eq:third-order-action-full}.

\begin{subequations}
	\begin{align*}
	\Gamma=\;&\mathcal{M}-2 \dot{\phi }^2 \mathcal{M}_X+\frac{3 \dot{\phi }^2 \mathcal{T}}{2},\numberthis\\
	\tilde\Xi=\;& \Xi+9 \Delta _0 \tilde{\Theta } +\frac{3}{2} \dot{\Delta }_0 \left(-8 \dot{\phi }^2 \mathcal{M}_X+3 \dot{\phi }^2 \mathcal{T}+4 \mathcal{M}+6\Delta_0\cM\right)-3 \dot{\Delta }_1 \mathcal{M},\numberthis\\
	\hat\Sigma=\;&\tilde{\Sigma }+\frac32 \dot{\phi }^2 \ddot{\Delta }_0 \mathcal{T}+3 \Delta _0 \ddot{\Delta }_0 \mathcal{M}+6 \dot{\Delta }_0{}^2 \mathcal{M}+\frac{3}{2}\dot{\Delta }_0 \left( 3 H \dot{\phi }^2 \mathcal{T}-2 \dot{\phi }^3 \ddot{\phi } \mathcal{T}_X\right.\\
	&\left.+\dot{\phi }^3 \mathcal{T}_{\phi }+2 \dot{\phi } \ddot{\phi } \mathcal{T}+2 \Delta _0 (3 H \mathcal{M}-2 \dot{\phi } \ddot{\phi } \mathcal{M}_X+\dot{\phi } \mathcal{M}_{\phi })\right)\numberthis\\
	\tilde\Omega=\;&\Omega+\Delta _0 \left(3 \dot{\Delta }_0 \left(3 \tilde{\Theta }-12 H \dot{\phi }^2 \mathcal{M}_X-9 H \dot{\phi }^2 \mathcal{T}+6 H \mathcal{M}-3 \dot{\phi }^3 \mathcal{T}_{\phi }-4 \dot{\phi }^3 \mathcal{M}_{\text{$\phi $X}}+2 \dot{\phi } \mathcal{M}_{\phi }\right)\right.\\
	&\left.-9 \tilde{\Sigma }+3 H \Xi +\dot{\Xi }+3 \ddot{\Delta }_0 \left(-4 \dot{\phi }^2 \mathcal{M}_X-3 \dot{\phi }^2 \mathcal{T}+2 \mathcal{M}\right)-45 \dot{\Delta }_0{}^2 \mathcal{M}\right)\\
	&+6 \Delta _1 \left(3 H \tilde{\Theta }+\dot{\tilde{\Theta }}\right)+\Delta _0^2 \left(-9 \left(3 H \tilde{\Theta }+\dot{\tilde{\Theta }}\right)-36 \dot{\Delta }_0 (3 H \mathcal{M}+\dot{\phi } \mathcal{M}_{\phi })-36 \ddot{\Delta }_0 \mathcal{M}\right)\\
	&-3 \dot{\Delta }_1 \tilde{\Theta }-3 \dot{\Delta }_0{}^2 \left(-2 \dot{\phi }^2 \mathcal{M}_X+3 \dot{\phi }^2 \mathcal{T}+\mathcal{M}\right)+\ddot{\Delta }_0 \left(6 \left(2 \dot{\phi }^2 \mathcal{T}-\dot{\phi }^4 \mathcal{T}_X\right)+6 \Delta _1 \mathcal{M}\right)\\
	&+\dot{\Delta }_0 \left(-18 H \dot{\phi }^4 \mathcal{T}_X+36 H \dot{\phi }^2 \mathcal{T}+6 \Delta _1 (3 H \mathcal{M}+\dot{\phi } \mathcal{M}_{\phi })-2 \Xi -6 \dot{\phi }^5 \mathcal{T}_{\text{$\phi $X}}+12 \dot{\phi }^3 \mathcal{T}_{\phi }-3 \dot{\Delta }_1 \mathcal{M}\right)\\
	&+\ddot{\phi } \left(-6 \Delta _0 \dot{\Delta }_0 \left(-3 \dot{\phi }^3 \mathcal{T}_X-4 \dot{\phi }^3 \mathcal{M}_{\text{XX}}+6 \dot{\phi } \mathcal{M}_X+3 \dot{\phi } \mathcal{T}\right)\right.\\
	&\left.+\dot{\Delta }_0 \left(12 \left(\dot{\phi }^5 \mathcal{T}_{\text{XX}}-4 \dot{\phi }^3 \mathcal{T}_X+2 \dot{\phi } \mathcal{T}\right)-12 \Delta _1 \dot{\phi } \mathcal{M}_X\right)+72 \Delta _0^2 \dot{\phi } \dot{\Delta }_0 \mathcal{M}_X\right),\numberthis\\
	\Omega=\;&-9 H^2 \mathcal{M}-2 \dot{\phi }^6 \left(9 H^2 \mathcal{T}_{XX}-6 H^2 \mathcal{M}_{XXX}+3 \dot{H} \mathcal{T}_{XX}+2 F_{XXX}\right)\\
	&-6 H \dot{\phi }^7 (\mathcal{T}_{\phi XX}-4 F_{2\phi XXX})+3 \dot{\phi }^4 \left(33 H^2 \mathcal{T}_X-24 H^2 \mathcal{M}_{XX}+11 \dot{H} \mathcal{T}_X+4 F_{XX}\right)\\
	&-3 \dot{\phi }^2 \left(-27 H^2 \mathcal{M}_X+30 H^2 \mathcal{T}+10 \dot{H} \mathcal{T}+F_X\right)+3 H \dot{\phi }^5 (11 \mathcal{T}_{\phi X}-48 F_{2\phi XX})\\
	&-6 H \dot{\phi }^3 (5 \mathcal{T}_{\phi }-27 F_{2\phi X})-18 H \dot{\phi } F_{2\phi }+\dddot{\phi } \left(-4 \dot{\phi }^5 \mathcal{B}_{XX}+22 \dot{\phi }^3 \mathcal{B}_X-20 \dot{\phi } \mathcal{B}\right)\\
	&+\ddot{\phi }^2 \left(4 \dot{\phi }^6 \mathcal{B}_{XXX}-28 \dot{\phi }^4 \mathcal{B}_{XX}+31 \dot{\phi }^2 \mathcal{B}_X+10 \mathcal{B}\right)\\
	&+\ddot{\phi } \left(-6 H \dot{\phi }^5 (2 \mathcal{B}_{XX}-\mathcal{T}_{XX})+33 H \dot{\phi }^3 (2 \mathcal{B}_X-\mathcal{T}_X)-30 H \dot{\phi } (2 \mathcal{B}-\mathcal{T})\right.\\
	&\left.-4 \dot{\phi }^6 \mathcal{B}_{\phi XX}+22 \dot{\phi }^4 \mathcal{B}_{\phi X}-20 \dot{\phi }^2 \mathcal{B}_{\phi }\right),\numberthis\\
	\Xi=\;&3 \dot{\phi }^4 (3 H \mathcal{T}_X-4 H \mathcal{M}_{XX})-6 \dot{\phi }^2 (3 H \mathcal{T}-5 H \mathcal{M}_X)-6 H \mathcal{M}-12 \dot{\phi }^5 F_{2\phi XX}+30 \dot{\phi }^3 F_{2\phi X}\\
	&+3\ddot{\phi }\dot\phi \left(2 \dot{\phi }^4 \mathcal{T}_{XX}+\dot{\phi }^2 (2 \mathcal{B}_X-9 \mathcal{T}_X)-2 (2 \mathcal{B}-3 \mathcal{T})\right)-6 \dot{\phi } F_{2\phi }.\numberthis
	\end{align*}
\end{subequations}

\bibliographystyle{JHEP3}
\bibliography{main}

\end{document}